\documentclass[10pt,journal,compsoc]{IEEEtran}
\usepackage{url}
\usepackage[utf8]{inputenc}
\usepackage{amsmath}
\usepackage{amssymb}
\usepackage{xspace}
\usepackage{epsfig}
\usepackage{balance}
\usepackage{listings}
\usepackage[dvipsnames]{xcolor}

\definecolor{codegray}{rgb}{0.25,0.25,0.25} 
\definecolor{codepurple}{rgb}{0.58,0,0.82}
\lstdefinestyle{mystyle-yaml}{
  commentstyle=\color{gray},
  keywordstyle=\color{purple},
  numberstyle=\tiny\color{codegray},
  stringstyle=\color{codepurple},
  basicstyle=\color{Periwinkle}\ttfamily\scriptsize,
  rulecolor=\color{black},
  breakatwhitespace=true,         
  breaklines=true,                 
  captionpos=b,
  frame=tb,
  keepspaces=true,                 
  numbers=left,                    
  numbersep=5pt,                  
  showspaces=false,                
  showstringspaces=false,
  showtabs=false,                  
  tabsize=2,
  xleftmargin=10pt,
  belowskip=-10pt,
}

\lstdefinelanguage{yaml}{
  alsoletter={-},
  keywords={true,false,null,y,n,-},
  sensitive=false,
  comment=[l]{\#},
  morecomment=[s]{/*}{*/},
  moredelim=[l][\color{orange}]{\&},
  moredelim=[l][\color{magenta}]{*},
  moredelim=**[il][\color{purple}{:}\color{MidnightBlue}]{:},   
  morestring=[b]',
  morestring=[b]",
}

\usepackage[acronyms,nonumberlist,nopostdot,nomain,nogroupskip,acronymlists={hidden}]{glossaries}
\newglossary[algh]{hidden}{acrh}{acnh}{Hidden Acronyms}

\usepackage{booktabs}
\usepackage{tabularx}

\usepackage{tikz}
\usepackage{pgfplots}
\pgfplotsset{compat=newest}
\pgfplotsset{plot coordinates/math parser=false}
\newlength\fheight
\newlength\fwidth
\usetikzlibrary{plotmarks,patterns,decorations.pathreplacing,backgrounds,calc,arrows,arrows.meta,spy,matrix,scopes}
\usepgfplotslibrary{patchplots,groupplots}
\usepackage{tikzscale}
\usepackage[draft]{hyperref}

\newif\ifexttikz
\exttikzfalse

\ifexttikz
	\usetikzlibrary{external}
\fi

\usepackage{multirow}
\usepackage[font=footnotesize]{subcaption}
\usepackage[font=footnotesize]{caption}

\usepackage{mathtools}
\usepackage[numbers,sort&compress]{natbib}

\usepackage{soul}

\newacronym{3gpp}{3GPP}{3rd Generation Partnership Project}
\newacronym{4g}{4G}{4th generation}
\newacronym{5g}{5G}{5th generation}
\newacronym{6g}{6G}{6th generation}
\newacronym{5gc}{5GC}{5G Core}
\newacronym{adc}{ADC}{Analog to Digital Converter}
\newacronym{aerpaw}{AERPAW}{Aerial Experimentation and Research Platform for Advanced Wireless}
\newacronym{ai}{AI}{Artificial Intelligence}
\newacronym{aimd}{AIMD}{Additive Increase Multiplicative Decrease}
\newacronym{am}{AM}{Acknowledged Mode}
\newacronym{amc}{AMC}{Adaptive Modulation and Coding}
\newacronym{amf}{AMF}{Access and Mobility Management Function}
\newacronym{aops}{AOPS}{Adaptive Order Prediction Scheduling}
\newacronym{api}{API}{Application Programming Interface}
\newacronym{apn}{APN}{Access Point Name}
\newacronym{ap}{AP}{Application Protocol}
\newacronym{aqm}{AQM}{Active Queue Management}
\newacronym{ausf}{AUSF}{Authentication Server Function}
\newacronym{avc}{AVC}{Advanced Video Coding}
\newacronym{awgn}{AGWN}{Additive White Gaussian Noise}
\newacronym{balia}{BALIA}{Balanced Link Adaptation Algorithm}
\newacronym{bbu}{BBU}{Base Band Unit}
\newacronym{bdp}{BDP}{Bandwidth-Delay Product}
\newacronym{ber}{BER}{Bit Error Rate}
\newacronym{bf}{BF}{Beamforming}
\newacronym{bler}{BLER}{Block Error Rate}
\newacronym{brr}{BRR}{Bayesian Ridge Regressor}
\newacronym{bs}{BS}{Base Station}
\newacronym{bsr}{BSR}{Buffer Status Report}
\newacronym{bss}{BSS}{Business Support System}
\newacronym{ca}{CA}{Carrier Aggregation}
\newacronym{caas}{CaaS}{Connectivity-as-a-Service}
\newacronym{cb}{CB}{Code Block}
\newacronym{cc}{CC}{Congestion Control}
\newacronym{ccid}{CCID}{Congestion Control ID}
\newacronym{cco}{CC}{Carrier Component}
\newacronym{cd}{CD}{Continuous Delivery}
\newacronym{cdd}{CDD}{Cyclic Delay Diversity}
\newacronym{cdf}{CDF}{Cumulative Distribution Function}
\newacronym{cdn}{CDN}{Content Distribution Network}
\newacronym{cli}{CLI}{Command-line Interface}
\newacronym{cn}{CN}{Core Network}
\newacronym{codel}{CoDel}{Controlled Delay Management}
\newacronym{comac}{COMAC}{Converged Multi-Access and Core}
\newacronym{cord}{CORD}{Central Office Re-architected as a Datacenter}
\newacronym{cornet}{CORNET}{COgnitive Radio NETwork}
\newacronym{cosmos}{COSMOS}{Cloud Enhanced Open Software Defined Mobile Wireless Testbed for City-Scale Deployment}
\newacronym{cots}{COTS}{Commercial Off-the-Shelf}
\newacronym{cp}{CP}{Control Plane}
\newacronym{cyp}{CP}{Cyclic Prefix}
\newacronym{up}{UP}{User Plane}
\newacronym{cpu}{CPU}{Central Processing Unit}
\newacronym{cqi}{CQI}{Channel Quality Information}
\newacronym{cr}{CR}{Cognitive Radio}
\newacronym{cran}{CRAN}{Cloud \gls{ran}}
\newacronym{crs}{CRS}{Cell Reference Signal}
\newacronym{csi}{CSI}{Channel State Information}
\newacronym{csirs}{CSI-RS}{Channel State Information - Reference Signal}
\newacronym{cu}{CU}{Central Unit}
\newacronym{d2tcp}{D$^2$TCP}{Deadline-aware Data center TCP}
\newacronym{d3}{D$^3$}{Deadline-Driven Delivery}
\newacronym{dac}{DAC}{Digital to Analog Converter}
\newacronym{dag}{DAG}{Directed Acyclic Graph}
\newacronym{das}{DAS}{Distributed Antenna System}
\newacronym{dash}{DASH}{Dynamic Adaptive Streaming over HTTP}
\newacronym{dc}{DC}{Dual Connectivity}
\newacronym{dccp}{DCCP}{Datagram Congestion Control Protocol}
\newacronym{dce}{DCE}{Direct Code Execution}
\newacronym{dci}{DCI}{Downlink Control Information}
\newacronym{dctcp}{DCTCP}{Data Center TCP}
\newacronym{dl}{DL}{Downlink}
\newacronym{dmr}{DMR}{Deadline Miss Ratio}
\newacronym{dmrs}{DMRS}{DeModulation Reference Signal}
\newacronym{drlcc}{DRL-CC}{Deep Reinforcement Learning Congestion Control}
\newacronym{drs}{DRS}{Discovery Reference Signal}
\newacronym{du}{DU}{Distributed Unit}
\newacronym{e2e}{E2E}{end-to-end}
\newacronym{earfcn}{EARFCN}{E-UTRA Absolute Radio Frequency Channel Number}
\newacronym{ecaas}{ECaaS}{Edge-Cloud-as-a-Service}
\newacronym{ecn}{ECN}{Explicit Congestion Notification}
\newacronym{edf}{EDF}{Earliest Deadline First}
\newacronym{embb}{eMBB}{Enhanced Mobile Broadband}
\newacronym{empower}{EMPOWER}{EMpowering transatlantic PlatfOrms for advanced WirEless Research}
\newacronym{enb}{eNB}{evolved Node Base}
\newacronym{endc}{EN-DC}{E-UTRAN-\gls{nr} \gls{dc}}
\newacronym{epc}{EPC}{Evolved Packet Core}
\newacronym{eps}{EPS}{Evolved Packet System}
\newacronym{es}{ES}{Edge Server}
\newacronym{etsi}{ETSI}{European Telecommunications Standards Institute}
\newacronym[firstplural=Estimated Times of Arrival (ETAs)]{eta}{ETA}{Estimated Time of Arrival}
\newacronym{eutran}{E-UTRAN}{Evolved Universal Terrestrial Access Network}
\newacronym{faas}{FaaS}{Function-as-a-Service}
\newacronym{fapi}{FAPI}{Functional Application Platform Interface}
\newacronym{fdd}{FDD}{Frequency Division Duplexing}
\newacronym{fdm}{FDM}{Frequency Division Multiplexing}
\newacronym{fdma}{FDMA}{Frequency Division Multiple Access}
\newacronym{fed4fire}{FED4FIRE+}{Federation 4 Future Internet Research and Experimentation Plus}
\newacronym{fir}{FIR}{Finite Impulse Response}
\newacronym{fit}{FIT}{Future \acrlong{iot}}
\newacronym{fpga}{FPGA}{Field Programmable Gate Array}
\newacronym{fr2}{FR2}{Frequency Range 2}
\newacronym{fs}{FS}{Fast Switching}
\newacronym{fscc}{FSCC}{Flow Sharing Congestion Control}
\newacronym{ftp}{FTP}{File Transfer Protocol}
\newacronym{fw}{FW}{Flow Window}
\newacronym{ge}{GE}{Gaussian Elimination}
\newacronym{gnb}{gNB}{Next Generation Node Base}
\newacronym{gop}{GOP}{Group of Pictures}
\newacronym{gpr}{GPR}{Gaussian Process Regressor}
\newacronym{gpu}{GPU}{Graphics Processing Unit}
\newacronym{gtp}{GTP}{GPRS Tunneling Protocol}
\newacronym{gtpc}{GTP-C}{GPRS Tunnelling Protocol Control Plane}
\newacronym{gtpu}{GTP-U}{GPRS Tunnelling Protocol User Plane}
\newacronym{gtpv2c}{GTPv2-C}{\gls{gtp} v2 - Control}
\newacronym{gw}{GW}{Gateway}
\newacronym{harq}{HARQ}{Hybrid Automatic Repeat reQuest}
\newacronym{hetnet}{HetNet}{Heterogeneous Network}
\newacronym{hh}{HH}{Hard Handover}
\newacronym{hol}{HOL}{Head-of-Line}
\newacronym{hqf}{HQF}{Highest-quality-first}
\newacronym{hss}{HSS}{Home Subscription Server}
\newacronym{http}{HTTP}{HyperText Transfer Protocol}
\newacronym{ia}{IA}{Initial Access}
\newacronym{iab}{IAB}{Integrated Access and Backhaul}
\newacronym{ic}{IC}{Incident Command}
\newacronym{ietf}{IETF}{Internet Engineering Task Force}
\newacronym{imsi}{IMSI}{International Mobile Subscriber Identity}
\newacronym{imt}{IMT}{International Mobile Telecommunication}
\newacronym{iot}{IoT}{Internet of Things}
\newacronym{ip}{IP}{Internet Protocol}
\newacronym{itu}{ITU}{International Telecommunication Union}
\newacronym{kpi}{KPI}{Key Performance Indicator}
\newacronym{kpm}{KPM}{Key Performance Measurement}
\newacronym{kvm}{KVM}{Kernel-based Virtual Machine}
\newacronym{los}{LOS}{Line-of-Sight}
\newacronym{lsm}{LSM}{Link-to-System Mapping}
\newacronym{lstm}{LSTM}{Long Short Term Memory}
\newacronym{lte}{LTE}{Long Term Evolution}
\newacronym{lxc}{LXC}{Linux Container}
\newacronym{m2m}{M2M}{Machine to Machine}
\newacronym{mac}{MAC}{Medium Access Control}
\newacronym{manet}{MANET}{Mobile Ad Hoc Network}
\newacronym{mano}{MANO}{Management and Orchestration}
\newacronym{mc}{MC}{Multi-Connectivity}
\newacronym{mcc}{MCC}{Mobile Cloud Computing}
\newacronym{mchem}{MCHEM}{Massive Channel Emulator}
\newacronym{mcs}{MCS}{Modulation and Coding Scheme}
\newacronym{mec}{MEC}{Multi-access Edge Computing}
\newacronym{mec2}{MEC}{Mobile Edge Cloud}
\newacronym{mfc}{MFC}{Mobile Fog Computing}
\newacronym{mgen}{MGEN}{Multi-Generator}
\newacronym{mi}{MI}{Mutual Information}
\newacronym{mib}{MIB}{Master Information Block}
\newacronym{miesm}{MIESM}{Mutual Information Based Effective SINR}
\newacronym{mimo}{MIMO}{Multiple Input, Multiple Output}
\newacronym{ml}{ML}{Machine Learning}
\newacronym{mlr}{MLR}{Maximum-local-rate}
\newacronym[plural=\gls{mme}s,firstplural=Mobility Management Entities (MMEs)]{mme}{MME}{Mobility Management Entity}
\newacronym{mmtc}{mMTC}{Massive Machine-Type Communications}
\newacronym{mmwave}{mmWave}{millimeter wave}
\newacronym{mpdccp}{MP-DCCP}{Multipath Datagram Congestion Control Protocol}
\newacronym{mptcp}{MPTCP}{Multipath TCP}
\newacronym{mr}{MR}{Maximum Rate}
\newacronym{mrdc}{MR-DC}{Multi \gls{rat} \gls{dc}}
\newacronym{mse}{MSE}{Mean Square Error}
\newacronym{mss}{MSS}{Maximum Segment Size}
\newacronym{mt}{MT}{Mobile Termination}
\newacronym{mtd}{MTD}{Machine-Type Device}
\newacronym{mtu}{MTU}{Maximum Transmission Unit}
\newacronym{mumimo}{MU-MIMO}{Multi-user \gls{mimo}}
\newacronym{mvno}{MVNO}{Mobile Virtual Network Operator}
\newacronym{nalu}{NALU}{Network Abstraction Layer Unit}
\newacronym{nas}{NAS}{Network Attached Storage}
\newacronym{nat}{NAT}{Network Address Translation}
\newacronym{nbiot}{NB-IoT}{Narrow Band IoT}
\newacronym{nfv}{NFV}{Network Function Virtualization}
\newacronym{nfvi}{NFVI}{Network Function Virtualization Infrastructure}
\newacronym{ni}{NI}{Network Interfaces}
\newacronym{nic}{NIC}{Network Interface Card}
\newacronym{nlos}{NLOS}{Non-Line-of-Sight}
\newacronym{now}{NOW}{Non Overlapping Window}
\newacronym{nsm}{NSM}{Network Service Mesh}
\newacronym[type=hidden]{nr}{NR}{New Radio}
\newacronym{nrf}{NRF}{Network Repository Function}
\newacronym{nsa}{NSA}{Non Stand Alone}
\newacronym{nse}{NSE}{Network Slicing Engine}
\newacronym{nssf}{NSSF}{Network Slice Selection Function}
\newacronym{o2i}{O2I}{Outdoor to Indoor}
\newacronym{oai}{OAI}{OpenAirInterface}
\newacronym{oaicn}{OAI-CN}{\gls{oai} \acrlong{cn}}
\newacronym{oairan}{OAI-RAN}{\acrlong{oai} \acrlong{ran}}
\newacronym{oam}{OAM}{Operations, Administration and Maintenance}
\newacronym{ofdm}{OFDM}{Orthogonal Frequency Division Multiplexing}
\newacronym{olia}{OLIA}{Opportunistic Linked Increase Algorithm}
\newacronym{omec}{OMEC}{Open Mobile Evolved Core}
\newacronym{onap}{ONAP}{Open Network Automation Platform}
\newacronym{onf}{ONF}{Open Networking Foundation}
\newacronym{onos}{ONOS}{Open Networking Operating System}
\newacronym{oom}{OOM}{\gls{onap} Operations Manager}
\newacronym{opnfv}{OPNFV}{Open Platform for \gls{nfv}}
\newacronym[type=hidden]{oran}{O-RAN}{Open \gls{ran}}
\newacronym{orbit}{ORBIT}{Open-Access Research Testbed for Next-Generation Wireless Networks}
\newacronym{os}{OS}{Operating System}
\newacronym{oss}{OSS}{Operations Support System}
\newacronym{pa}{PA}{Position-aware}
\newacronym{pase}{PASE}{Prioritization, Arbitration, and Self-adjusting Endpoints}
\newacronym{pawr}{PAWR}{Platforms for Advanced Wireless Research}
\newacronym{pbch}{PBCH}{Physical Broadcast Channel}
\newacronym{pcef}{PCEF}{Policy and Charging Enforcement Function}
\newacronym{pcfich}{PCFICH}{Physical Control Format Indicator Channel}
\newacronym{pcrf}{PCRF}{Policy and Charging Rules Function}
\newacronym{pdcch}{PDCCH}{Physical Downlink Control Channel}
\newacronym{pdcp}{PDCP}{Packet Data Convergence Protocol}
\newacronym{pdf}{PDF}{Probability Density Function}
\newacronym{pdsch}{PDSCH}{Physical Downlink Shared Channel}
\newacronym{pdu}{PDU}{Packet Data Unit}
\newacronym{pf}{PF}{Proportional Fair}
\newacronym{pgw}{PGW}{Packet Gateway}
\newacronym{phich}{PHICH}{Physical Hybrid ARQ Indicator Channel}
\newacronym{phy}{PHY}{Physical}
\newacronym{pmch}{PMCH}{Physical Multicast Channel}
\newacronym{pmi}{PMI}{Precoding Matrix Indicators}
\newacronym{powder}{POWDER}{Platform for Open Wireless Data-driven Experimental Research}
\newacronym{ppo}{PPO}{Proximal Policy Optimization}
\newacronym{ppp}{PPP}{Poisson Point Process}
\newacronym{prach}{PRACH}{Physical Random Access Channel}
\newacronym{prb}{PRB}{Physical Resource Block}
\newacronym{psnr}{PSNR}{Peak Signal to Noise Ratio}
\newacronym{pss}{PSS}{Primary Synchronization Signal}
\newacronym{pucch}{PUCCH}{Physical Uplink Control Channel}
\newacronym{pusch}{PUSCH}{Physical Uplink Shared Channel}
\newacronym{qam}{QAM}{Quadrature Amplitude Modulation}
\newacronym{qci}{QCI}{\gls{qos} Class Identifier}
\newacronym{qoe}{QoE}{Quality of Experience}
\newacronym{qos}{QoS}{Quality of Service}
\newacronym{quic}{QUIC}{Quick UDP Internet Connections}
\newacronym{rach}{RACH}{Random Access Channel}
\newacronym{ran}{RAN}{Radio Access Network}
\newacronym[firstplural=Radio Access Technologies (RATs)]{rat}{RAT}{Radio Access Technology}
\newacronym{rbg}{RBG}{Resource Block Group}
\newacronym{rcn}{RCN}{Research Coordination Network}
\newacronym{rc}{RC}{RAN Control}
\newacronym{rec}{REC}{Radio Edge Cloud}
\newacronym{red}{RED}{Random Early Detection}
\newacronym{renew}{RENEW}{Reconfigurable Eco-system for Next-generation End-to-end Wireless}
\newacronym{rf}{RF}{Radio Frequency}
\newacronym{rfc}{RFC}{Request for Comments}
\newacronym{rfr}{RFR}{Random Forest Regressor}
\newacronym{ric}{RIC}{\gls{ran} Intelligent Controller}
\newacronym{rlc}{RLC}{Radio Link Control}
\newacronym{rlf}{RLF}{Radio Link Failure}
\newacronym{rlnc}{RLNC}{Random Linear Network Coding}
\newacronym{rmr}{RMR}{RIC Message Router}
\newacronym{rmse}{RMSE}{Root Mean Squared Error}
\newacronym{rnis}{RNIS}{Radio Network Information Service}
\newacronym{rr}{RR}{Round Robin}
\newacronym{rrc}{RRC}{Radio Resource Control}
\newacronym{rrm}{RRM}{Radio Resource Management}
\newacronym{rru}{RRU}{Remote Radio Unit}
\newacronym{rs}{RS}{Remote Server}
\newacronym{rsrp}{RSRP}{Reference Signal Received Power}
\newacronym{rsrq}{RSRQ}{Reference Signal Received Quality}
\newacronym{rss}{RSS}{Received Signal Strength}
\newacronym{rssi}{RSSI}{Received Signal Strength Indicator}
\newacronym{rtt}{RTT}{Round Trip Time}
\newacronym{ru}{RU}{Radio Unit}
\newacronym{rw}{RW}{Receive Window}
\newacronym{rx}{RX}{Receiver}
\newacronym{s1ap}{S1AP}{S1 Application Protocol}
\newacronym{sa}{SA}{standalone}
\newacronym{sack}{SACK}{Selective Acknowledgment}
\newacronym{sap}{SAP}{Service Access Point}
\newacronym{sc2}{SC2}{Spectrum Collaboration Challenge}
\newacronym{scef}{SCEF}{Service Capability Exposure Function}
\newacronym{sch}{SCH}{Secondary Cell Handover}
\newacronym{scoot}{SCOOT}{Split Cycle Offset Optimization Technique}
\newacronym{sctp}{SCTP}{Stream Control Transmission Protocol}
\newacronym{sdap}{SDAP}{Service Data Adaptation Protocol}
\newacronym{sdk}{SDK}{Software Development Kit}
\newacronym{sdm}{SDM}{Space Division Multiplexing}
\newacronym{sdma}{SDMA}{Spatial Division Multiple Access}
\newacronym{sdn}{SDN}{Software-defined Networking}
\newacronym{sdr}{SDR}{Software-defined Radio}
\newacronym{seba}{SEBA}{SDN-Enabled Broadband Access}
\newacronym{sgsn}{SGSN}{Serving GPRS Support Node}
\newacronym{sgw}{SGW}{Service Gateway}
\newacronym{si}{SI}{Study Item}
\newacronym{sib}{SIB}{Secondary Information Block}
\newacronym{sinr}{SINR}{Signal to Interference plus Noise Ratio}
\newacronym{sip}{SIP}{Session Initiation Protocol}
\newacronym{siso}{SISO}{Single Input, Single Output}
\newacronym{sla}{SLA}{Service Level Agreement}
\newacronym{sm}{SM}{Service Model}
\newacronym{smf}{SMF}{Session Management Function}
\newacronym{smo}{SMO}{Service Management and Orchestration}
\newacronym{sms}{SMS}{Short Message Service}
\newacronym{smsgmsc}{SMS-GMSC}{\gls{sms}-Gateway}
\newacronym{snr}{SNR}{Signal-to-Noise-Ratio}
\newacronym{son}{SON}{Self-Organizing Network}
\newacronym{sptcp}{SPTCP}{Single Path TCP}
\newacronym{srb}{SRB}{Service Radio Bearer}
\newacronym{srn}{SRN}{Standard Radio Node}
\newacronym{srs}{SRS}{Sounding Reference Signal}
\newacronym{ss}{SS}{Synchronization Signal}
\newacronym{sss}{SSS}{Secondary Synchronization Signal}
\newacronym{st}{ST}{Spanning Tree}
\newacronym{svc}{SVC}{Scalable Video Coding}
\newacronym{tb}{TB}{Transport Block}
\newacronym{tcp}{TCP}{Transmission Control Protocol}
\newacronym{tdd}{TDD}{Time Division Duplexing}
\newacronym{tdm}{TDM}{Time Division Multiplexing}
\newacronym{tdma}{TDMA}{Time Division Multiple Access}
\newacronym{tfl}{TfL}{Transport for London}
\newacronym{tfrc}{TFRC}{TCP-Friendly Rate Control}
\newacronym{tft}{TFT}{Traffic Flow Template}
\newacronym{tgen}{TGEN}{Traffic Generator}
\newacronym{tip}{TIP}{Telecom Infra Project}
\newacronym{tm}{TM}{Transparent Mode}
\newacronym{to}{TO}{Telco Operator}
\newacronym{tr}{TR}{Technical Report}
\newacronym{trp}{TRP}{Transmitter Receiver Pair}
\newacronym{ts}{TS}{Technical Specification}
\newacronym{tti}{TTI}{Transmission Time Interval}
\newacronym{ttt}{TTT}{Time-to-Trigger}
\newacronym{tx}{TX}{Transmitter}
\newacronym{uas}{UAS}{Unmanned Aerial System}
\newacronym{uav}{UAV}{Unmanned Aerial Vehicle}
\newacronym{udm}{UDM}{Unified Data Management}
\newacronym{udp}{UDP}{User Datagram Protocol}
\newacronym{udr}{UDR}{Unified Data Repository}
\newacronym{ue}{UE}{User Equipment}
\newacronym{uhd}{UHD}{\gls{usrp} Hardware Driver}
\newacronym{ul}{UL}{Uplink}
\newacronym{um}{UM}{Unacknowledged Mode}
\newacronym{uml}{UML}{Unified Modeling Language}
\newacronym{upa}{UPA}{Uniform Planar Array}
\newacronym{upf}{UPF}{User Plane Function}
\newacronym{urllc}{URLLC}{Ultra Reliable and Low Latency Communications}
\newacronym{usa}{U.S.}{United States}
\newacronym{usim}{USIM}{Universal Subscriber Identity Module}
\newacronym{usrp}{USRP}{Universal Software Radio Peripheral}
\newacronym{utc}{UTC}{Urban Traffic Control}
\newacronym{vim}{VIM}{Virtualization Infrastructure Manager}
\newacronym{vm}{VM}{Virtual Machine}
\newacronym{vnf}{VNF}{Virtual Network Function}
\newacronym{volte}{VoLTE}{Voice over \gls{lte}}
\newacronym{voltha}{VOLTHA}{Virtual OLT HArdware Abstraction}
\newacronym{vr}{VR}{Virtual Reality}
\newacronym{vran}{vRAN}{Virtualized \gls{ran}}
\newacronym{vss}{VSS}{Video Streaming Server}
\newacronym{wbf}{WBF}{Wired Bias Function}
\newacronym{wf}{WF}{Waterfilling}
\newacronym{wg}{WG}{Working Group}
\newacronym{wlan}{WLAN}{Wireless Local Area Network}
\newacronym{osm}{OSM}{Open Source \gls{nfv} Management and Orchestration}
\newacronym{pnf}{PNF}{Physical Network Function}
\newacronym{drl}{DRL}{Deep Reinforcement Learning}
\newacronym{mtc}{MTC}{Machine-type Communications}
\newacronym{osc}{OSC}{O-RAN Software Community}
\newacronym{mns}{MnS}{Management Services}
\newacronym{ves}{VES}{\gls{vnf} Event Stream}
\newacronym{ei}{EI}{Enrichment Information}
\newacronym{fh}{FH}{Fronthaul}
\newacronym{fft}{FFT}{Fast Fourier Transform}
\newacronym{laa}{LAA}{Licensed-Assisted Access}
\newacronym{plfs}{PLFS}{Physical Layer Frequency Signals}
\newacronym{ptp}{PTP}{Precision Time Protocol}
\newacronym{cbrs}{CBRS}{Citizen Broadband Radio Service}
\tikzstyle{startstop} = [rectangle, rounded corners, minimum width=2cm, minimum height=0.5cm,text centered, draw=black]
\tikzstyle{io} = [trapezium, trapezium left angle=70, trapezium right angle=110, minimum width=3cm, minimum height=1cm, text centered, draw=black]
\tikzstyle{process} = [rectangle, minimum width=2cm, minimum height=0.5cm, text centered, draw=black, alignb=center]
\tikzstyle{decision} = [ellipse, minimum width=2cm, minimum height=1cm, text centered, draw=black]
\tikzstyle{arrow} = [thick,<->,>=stealth]
\tikzstyle{line} = [thick,>=stealth]
\tikzstyle{darrow} = [thick,<->,>=stealth,dashed]
\tikzstyle{sarrow} = [thick,->,>=stealth]
\tikzstyle{larrow} = [line width=0.1mm,dashdotted,->,>=stealth]
\tikzstyle{llarrow} = [line width=0.1mm,->,>=stealth]

\makeatletter
\def\grd@save@target#1{%
  \def\grd@target{#1}}
\def\grd@save@start#1{%
  \def\grd@start{#1}}
\tikzset{
  grid with coordinates/.style={
    to path={%
      \pgfextra{%
        \edef\grd@@target{(\tikztotarget)}%
        \tikz@scan@one@point\grd@save@target\grd@@target\relax
        \edef\grd@@start{(\tikztostart)}%
        \tikz@scan@one@point\grd@save@start\grd@@start\relax
        \draw[minor help lines] (\tikztostart) grid (\tikztotarget);
        \draw[major help lines] (\tikztostart) grid (\tikztotarget);
        \grd@start
        \pgfmathsetmacro{\grd@xa}{\the\pgf@x/1cm}
        \pgfmathsetmacro{\grd@ya}{\the\pgf@y/1cm}
        \grd@target
        \pgfmathsetmacro{\grd@xb}{\the\pgf@x/1cm}
        \pgfmathsetmacro{\grd@yb}{\the\pgf@y/1cm}
        \pgfmathsetmacro{\grd@xc}{\grd@xa + \pgfkeysvalueof{/tikz/grid with coordinates/major step x}}
        \pgfmathsetmacro{\grd@yc}{\grd@ya + \pgfkeysvalueof{/tikz/grid with coordinates/major step y}}
        \foreach \x in {\grd@xa,\grd@xc,...,\grd@xb}
        \node[anchor=north] at (\x,\grd@ya) {\pgfmathprintnumber{\x}};
        \foreach \y in {\grd@ya,\grd@yc,...,\grd@yb}
        \node[anchor=east] at (\grd@xa,\y) {\pgfmathprintnumber{\y}};
      }
    }
  },
  minor help lines/.style={
    help lines,
    gray,
    line cap =round,
    xstep=\pgfkeysvalueof{/tikz/grid with coordinates/minor step x},
    ystep=\pgfkeysvalueof{/tikz/grid with coordinates/minor step y}
  },
  major help lines/.style={
    help lines,
    line cap =round,
    line width=\pgfkeysvalueof{/tikz/grid with coordinates/major line width},
    xstep=\pgfkeysvalueof{/tikz/grid with coordinates/major step x},
    ystep=\pgfkeysvalueof{/tikz/grid with coordinates/major step y}
  },
  grid with coordinates/.cd,
  minor step x/.initial=.5,
  minor step y/.initial=.2,
  major step x/.initial=1,
  major step y/.initial=1,
  major line width/.initial=1pt,
}
\makeatother

\definecolor{desireRed}{RGB}{230,57,60}%
\definecolor{darkPurple}{RGB}{59,31,43}%
\definecolor{springGreen}{RGB}{37,223,145}%
\definecolor{queenBlue}{RGB}{69,123,157}%
\definecolor{spaceCadet}{RGB}{29,53,87}%

\usepackage{dblfloatfix}

\newcommand{\openrangym}{OpenRAN Gym\xspace}
\newcommand{\scope}{SCOPE\xspace}
\newcommand{\neutran}{NeutRAN\xspace}

\newcommand{\mf}{\mathbf}
\newcommand{\mc}{\mathcal}

\newcommand{\mr}{\mathrm}

\newcommand{\maximize}{\mathrm{maximize}}

\newcommand{\reqs}{\mathcal{I}}
\newcommand{\bss}{\mathcal{R}}
\newcommand{\freqs}{\mathcal{F}}
\newcommand{\tenants}{\mathcal{T}}
\newcommand{\bands}{\mathcal{W}}
\newcommand{\ran}{\gls{ran}\xspace}
\newcommand{\ric}{\gls{ric}\xspace}

\newcommand{\rus}{\glspl{ru}\xspace}

\newcommand{\areas}{\mc{A}}
\newcommand{\prb}{\gls{prb}\xspace}
\newcommand{\prbs}{\glspl{prb}\xspace}

\usepackage[var0]{inconsolata}

\begin{document}


\title{\neutran: An Open RAN Neutral Host Architecture for Zero-Touch RAN and\\Spectrum Sharing}

\author{\IEEEauthorblockN{%
Leonardo Bonati,~\IEEEmembership{Member, IEEE},
Michele Polese,~\IEEEmembership{Member, IEEE},
Salvatore D'Oro,~\IEEEmembership{Member, IEEE},
Stefano Basagni,~\IEEEmembership{Senior Member, IEEE},
Tommaso Melodia,~\IEEEmembership{Fellow, IEEE}}
\thanks{The authors are with the Institute for the Wireless Internet of Things, Northeastern University, Boston, MA, USA. E-mail: \{l.bonati, m.polese, s.doro, s.basagni, melodia\}@northeastern.edu.}
\thanks{This article is based upon work partially supported by the U.S.\ National Science Foundation under grants CNS-1925601, CNS-2112471, and CNS-1923789, by the U.S.\ Office of Naval Research under grant N00014-20-1-2132, and by OUSD(R\&E) through Army Research Laboratory Cooperative Agreement Number W911NF-19-2-0221. The views and conclusions contained in this document are those of the authors and should not be interpreted as representing the official policies, either expressed or implied, of the Army Research Laboratory or the U.S. Government. The U.S. Government is authorized to reproduce and distribute reprints for Government purposes notwithstanding any copyright notation herein.}
}

\flushbottom
\setlength{\parskip}{0ex plus0.1ex}

\maketitle
\glsunset{nr}
\glsunset{lte}
\glsunset{3gpp}
\glsunset{cbrs}

\begin{abstract}
Obtaining access to exclusive spectrum, cell sites, \ran equipment, and edge infrastructure imposes major capital expenses to mobile network operators. 
%
%
A neutral host infrastructure, by which a third-party company provides \gls{ran} services to mobile operators through network virtualization and slicing techniques, is seen as a promising solution to decrease these costs. 
Currently, however, neutral host providers lack automated and virtualized pipelines for onboarding new tenants and to provide elastic and on-demand allocation of resources 
matching operators' requirements.
%
To address this gap,
this paper presents \neutran, a zero-touch framework based on the O-RAN architecture to support applications on neutral hosts and automatic operator onboarding.
\neutran builds upon two key components: (i)~an optimization engine to guarantee coverage and to meet quality of service requirements while accounting for the limited amount of shared spectrum and \gls{ran} nodes, and (ii)~a fully virtualized and automated infrastructure that converts the output of the optimization engine into deployable micro-services to be executed at \gls{ran} nodes and cell sites. 
\neutran was prototyped on an OpenShift cluster and on a programmable testbed with~4 base stations and~10 users from~3 different tenants. 
We evaluate its benefits, comparing it to a traditional license-based \ran where each tenant has dedicated physical and spectrum resources.
We show that \neutran can deploy a fully operational neutral host-based cellular network in around~10 seconds. 
Experimental results show that it increases the cumulative network throughput by~2.18$\times$ and the per-user average throughput by~1.73$\times$
in networks with shared spectrum blocks of 30\:MHz. 
\neutran provides a~1.77$\times$ cumulative throughput gain even when it can only operate on a shared spectrum block of 10\:MHz (one third of the spectrum used in license-based RANs).
\end{abstract}


\begin{picture}(0,0)(10,-590)
\put(0,0){
\put(0,0){\footnotesize This paper has been accepted for publication on IEEE Transactions on Mobile Computing.}
\put(0,-10){
\scriptsize \textcopyright~2023 IEEE. Personal use of this material is permitted. Permission from IEEE must be obtained for all other uses, in any current or future media, including}
\put(0, -17){
\scriptsize reprinting/republishing this material for advertising or promotional purposes, creating new collective works, for resale or redistribution to servers or lists,}
\put(0, -24){
\scriptsize or reuse of any copyrighted component of this work in other works.}
}
\end{picture}

\glsresetall
\glsunset{nr}
\glsunset{lte}
\glsunset{3gpp}
\glsunset{cbrs}

\section{Introduction}
The need for higher data rates and reduced latency in cellular networks has resulted in unprecedented network densification~\cite{bhushan2014network}, and in new deployment models where private operators deploy dedicated cellular infrastructure~\cite{wen2022private}. As a consequence, access to spectrum, cell site facilities (i.e., poles, towers), \ran and edge equipment~\cite{fccReport} represents the largest share of capital and operational expenses faced by public and private operators~\cite{oughton2018cost}. 

However, higher costs---which frequently come with lower profits---are barriers to technological innovation for the future of cellular networks.
To lower these barriers, renting a \emph{neutral host infrastructure} from a third-party company that leases physical resources (e.g., spectrum, towers, \ran) to multiple operators on a shared-tenant basis is seen as a promising solution as it enables resource sharing and decreases the overall infrastructure costs~\cite{samdanis2016network}. 
Examples of commercial neutral host deployments are reported in several analyses and reports~\cite{LAHTEENMAKI2021102201,alpha2021} and, according to~\cite{marketresearch}, there is evidence showing that the neutral host market will reach \$9.56 billions by 2028.
%
%
Similar to what is happening in the infrastructure domain, \emph{spectrum sharing} has also been identified as an effective way to increase the overall spectral utilization~\cite{zhang2017survey}. 
In fact, recent estimates indicate that joint adoption of neutral host models and spectrum sharing techniques can potentially lead to savings of at least~30\% on network operational costs in the next five years~\cite{analysismason}.

\ran and spectrum sharing, however, are not ready for prime time in multi-operator network deployments~\cite{stl2022open}, especially because they still lack mechanisms to enable (i)~\emph{fine-grained sharing}, with multiple tenants sharing compute and spectrum slices from the same physical infrastructure, and (ii)~\emph{dynamic sharing} that allows infrastructure owners 
to fully leverage the statistical multiplexing of \ran and spectrum resources, and 
to tailor infrastructure parameters to tenant requirements that may change in a matter of seconds.
For example, spectrum sharing in the \gls{cbrs} band operates over time scales in the order of minutes~\cite{palola2017field}, limiting the flexibility of the system and spectrum utilization efficiency.   

Obstacles to further progress are both technological and strategic in nature, and include the following.

\noindent$\bullet$~\emph{Lack of automated and virtualized pipelines for multi-tenant management}.
Zero-touch, resilient, fault-tolerant automation frameworks are currently unavailable in \ran environments. 
These functionalities are necessary to ensure reliability and proper coordination among multiple tenants that dynamically share infrastructure and spectrum, without manual intervention and over-provisioning~\cite{garciaaviles2021nuberu}.

\noindent$\bullet$~\emph{Lack of timely management of the life cycle of network services}.
The dynamic allocation of spectrum and \ran infrastructure resources in a timely fashion is still a challenge, considering that complex software services, such as softwarized 5G \glspl{gnb}, need to be instantiated in a matter of seconds.
%
This is because of the lack of low-latency end-to-end pipelines that interface with, keep track of, and coordinate available \ran elements and resources, and that manage the life cycle of network services~\cite{liyanage2022survey,benzaid2020ai,doro2022orchestran}.

\noindent$\bullet$~\emph{Operators' perception of resource sharing as a risk}.
Due to the absence of reliable sharing solutions that can support \glspl{sla} through dynamic, fine-grained resource allocation provided through optimization engines~\cite{sharma2017dynamic}, operators perceive spectrum sharing as a risk~\cite{gsma2021sharing} and, therefore, prefer exclusive spectrum licensing.


\subsection{Novelty and Contribution}

This paper addresses these challenges and takes a fundamental step toward enabling \emph{zero-touch dynamic and fine-grained \ran and spectrum sharing} by introducing \neutran. 
%
%
\neutran is a neutral host framework 
that automatically manages the deployment of services on shared \ran and spectrum resources, based on high-level intents and requests from multiple tenants. 
The core innovation of \neutran is the design and prototyping of end-to-end pipelines that combine (i) a virtualized and automated \ran infrastructure with (ii) an optimization engine that takes decisions regarding \ran and spectrum sharing policies. 
The \neutran framework, which we develop on top of the O-RAN architecture, shows for the first time how virtualization and automation can be extended to a multi-tenant \ran, providing a fully managed and effective solution for private and public neutral host-based deployments.
%
This allows \neutran to break the traditional, isolated spectrum and infrastructure silos and to bring {\em dynamic, fine-grained statistical multiplexing to the \ran and spectrum}. 

\noindent
The main contributions of this work include the following.

\noindent \emph{1) \neutran zero-touch framework and automation pipelines.} 
We define, develop, and prototype \neutran over state-of-the-art tools for future cellular network innovation, which include OpenShift, Kubernetes, O-RAN and \glspl{sdr}.
As an open and virtualized framework, \neutran enables the deployment of complex, customized core and \ran micro-services in a matter of seconds 
(e.g.,~$9.55$\:s for a \gls{gnb}) 
from a centralized \gls{smo} entity to edge datacenters and cell sites that are part of the O-RAN O-Cloud. 
\neutran manages~5G disaggregated \glspl{gnb} as well as the deployment of O-RAN \glspl{ric} and their custom logic units
to satisfy tenant requests.
These logic units, namely xApps if deployed on the near-real-time (near-RT) \ric and rApps on the non-real-time (non-RT) \ric, get run-time \glspl{kpm} from the \gls{ran} nodes, and can control their functionalities through data-driven techniques acting at different time scales~\cite{bonati2021intelligence}.
%


\noindent \emph{2) \neutran optimization engine rApp.} 
We develop an optimization engine that provides guarantees for (i) the execution of latency-critical compute tasks (e.g., a \gls{gnb}) in shared infrastructures, and for (ii) the \gls{qos} and \glspl{sla} that tenants require for their users. 
The engine is based on the efficient solution of the \emph{neutral host problem}, which considers tenant requests, available resources, and network analytics 
to generate an optimal allocation of micro-services and spectrum resources (e.g., spectrum slices). 
The problem is modeled as a binary Quadratically Constrained Quadratic Programming (QCQP) optimization problem that is solved optimally via reformulation-linearization techniques. 
The \neutran optimization engine is deployed as an rApp in the \gls{smo}, and leverages O-RAN interfaces to gather data and analytics on the \ran performance and to deploy the \neutran services.

\noindent \emph{3) Scalability, efficiency, and experimental evaluation.}
%
We show that \neutran computes optimal solutions in less than~$2$\:s for large-scale networks, allocating resources that meet the tenant requirements. 
We also run experiments
on 
a testbed with 4~softwarized
cell sites and 10 commercial \glspl{ue} from 3~different tenants. 
We compare \neutran against a license-based, siloed \ran where operators control their own infrastructure and spectrum ($10$\:MHz each; $30$\:MHz total). 
Our results show that \neutran manages the total (now shared) bandwidth obtaining \emph{2.18$\times$ \ran throughput} and \emph{1.73$\times$ average user throughput gains}, with a consistent improvement in \gls{sinr} over the license-based approach.
%
We also show that, when \neutran is deployed on a total and shared bandwidth that is one third of the previous configurations ($10$\:MHz), the optimized resource allocation at cell sites offsets the reduction in available spectrum, delivering an improvement in \ran throughput of 1.77$\times$ and unchanged average user throughput.
This shows how the combination of virtualization, automation, and optimization to manage \ran and spectrum sharing brings remarkable network and user throughput gains.


The remainder of this paper is organized as follows. 
Section~\ref{sec:related} reviews the state of the art on neutral host applications, \ran and spectrum sharing. Section~\ref{sec:overview} presents \neutran. Section~\ref{sec:formulation} illustrates the neutral host problem and presents \neutran optimization engine.
Section~\ref{sec:prototype} describes the \neutran prototype implementation, and Section~\ref{sec:evaluation} profiles \neutran performance both via numerical and experimental results.
Finally, Section~\ref{sec:conclusions} concludes the paper.

\section{Related Work}
\label{sec:related}

The \neutran concept lies at the intersection of solutions for neutral host architectures, spectrum and \gls{ran} sharing, and controllers for the virtualized \gls{ran}, all fields that have garnered the interest of researchers from industry and academia alike.

%
Considering neutral host and virtualized architectures for \gls{ran} and spectrum sharing,
Kibria et al.\ study the business models enabled by micro-operators sharing neutral host architectures~\cite{kibria2017shared},
while Di Pascale et al.\ leverage the blockchain to enforce \glspl{sla} for
neutral host infrastructure providers~\cite{dipascale2020toward}.
Incentives for cooperation among operators are analyzed by Vincenzi et al.~\cite{vincenzi2017cooperation}.
Architectural enhancements for neutral hosts are presented by Sarakis et al.~\cite{sarakis2021cost} (on cost-efficient private networks) and by Paolino et al.~\cite{paolino2019compute} (on security for network virtualization).
%
%
%
Despite considering both neutral host and virtualization paradigms, the above works are either focused on the business aspects of such architectures, on cooperation incentives between tenants, or on architectural enhancements.

Multi-operator solutions for spectrum and \gls{ran} sharing have also been thoroughly investigated.
%
Sharing for the coexistence of \gls{iot} devices and cellular services is investigated by Xiao et al.~\cite{xiao2019multi}, and Qian et al.~\cite{qian2020multi}.
Lin et al.\ use \gls{ran} proxies to share base stations among
core networks~\cite{lin2017transparent}.
Giannone et al.\ 
focus on the deployment of virtualized \glspl{du}~\cite{giannone2019impact}.
%
Samdanis et al.\ propose a mechanism for tenants to lease resources from infrastructure providers~\cite{samdanis2016network}.
Kasgari and Saad design an optimization and control framework for isolation of multiple slices and for the reduction of their power consumption~\cite{kasgari2018stochastic}.
Wang et al.\ propose a hybrid slice reconfiguration framework to optimize allocated resources with a focus on the profit of each slice~\cite{wang2019reconfiguration}.
Bega et al.\ design a \gls{ml} framework for the reallocation of slice resources and revenue maximization based on the forecast network capacity in~\cite{bega2019deepcog}.
%
Dynamic slice sharing is evaluated by Foukas et al.\ in~\cite{foukas2019iris}, by Caballero et al.\ in~\cite{caballero2019network,caballero2017multi}, and by D'Oro et al. in~\cite{doro2021coordinated}, who also evaluate the sharing of compute and storage resources in~\cite{doro2020sledge}.
Puligheddu et al.\ propose an O-RAN-based semantic framework to allocate resources
to \gls{ml} classifiers based on the requirements of the task to perform at the edge~\cite{puligheddu2023semoran}; their focus is on computer vision applications.
%
Baldesi et al.\ propose a mechanism for dynamic adaptation based on incumbents in the shared spectrum~\cite{baldesi2022charm}.
While focusing on sharing network resources, these works propose solutions that are either not tailored to large deployments, do not adapt to network dynamics, or do not consider the automated deployment of micro-services, all functionalities offered by \neutran. Similarly, several papers focus on admission control for slicing~\cite{bega2017optimising,sciancalepore2017traffic,caballero2018netowrk}, but do not consider the dynamics and ad hoc deployment of \gls{ran} base stations as micro-services.

Controllers for virtualized \glspl{ran} and abstractions among the physical and virtual infrastructure are discussed by Schmidt et al.~\cite{schmidt2019flexvran}.
%
%
Moorthy et al.\ study the virtualization of control-plane functionalities~\cite{moorthy2022oswireless}.
%
Foukas et al.\ devise a slicing system for the dynamic virtualization of base stations~\cite{foukas2017orion} and showcase a neutral-host use case in ~\cite{neutral2018}.
Garcia-Aviles et al.\ focus on
synchronization issues
among shared \glspl{du}
and their \glspl{ue} in case of lack of computational capacity~\cite{garciaaviles2021nuberu}.
Even though these solutions provide enhanced and automated network control, they either do not focus on the automated instantiation of \ran functions as micro-services, or do not tackle neutral host architectures specifically, or focus on RAN slicing applications that do not consider the availability of multiple spectrum bands.

Compared to the listed works, \neutran proposes an end-to-end solution for the zero-touch automated and rApp-based optimized allocation and deployment of micro-services in a neutral host architecture based on tenant intents.
This is achieved by computing the optimized allocation of \ran functions
starting from intents, and by automatically instantiating such functions as micro-services on edge datacenters managed by enterprise-ready platform-as-a-service frameworks such as OpenShift.

\section{The \neutran Framework}
\label{sec:overview}



\begin{figure}[t]
    \centering
    \vspace{-.1cm}
    \includegraphics[width=0.95\columnwidth]{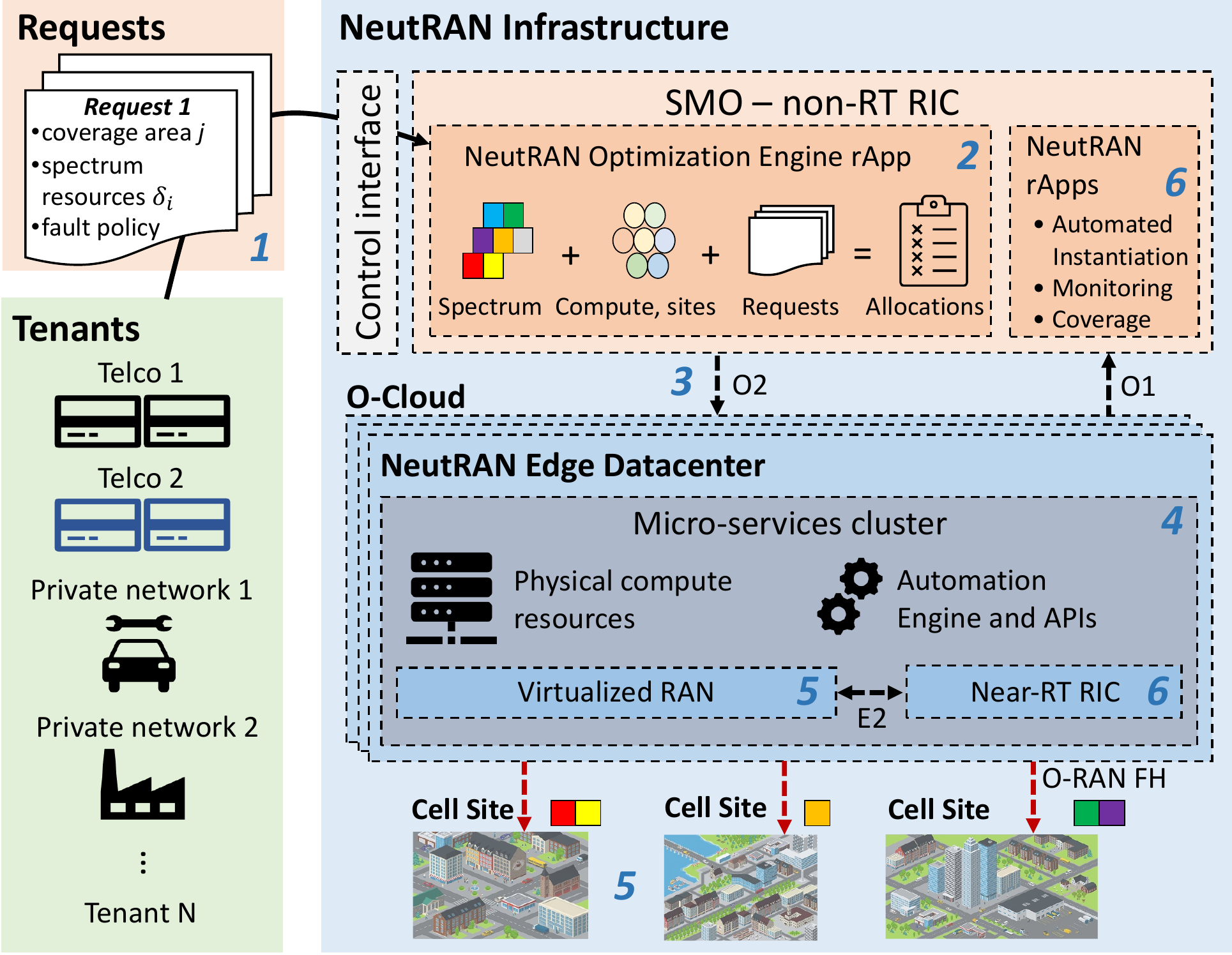}
    \setlength\belowcaptionskip{-.3cm}
    \caption{The \neutran framework. The numbering represents the six steps of the \neutran workflow.}
    \label{fig:architecture}
\end{figure}

\neutran combines an O-RAN-based softwarized and automated infrastructure and an optimization engine for practical and efficient \ran and spectrum sharing. A bird's-eye view of the \neutran framework is shown in Figure~\ref{fig:architecture}.

The \neutran stakeholders are the tenants, who want to provide services to their end users, and the \neutran operator, who owns the infrastructure and provides access to the automated \ran and spectrum sharing pipelines. 
Tenants access a high-level control interface to submit requests to deploy cellular connectivity in certain areas. Based on the available physical resources, these requests are then automatically converted into a set of virtualized networking services and functionalities deployed by \neutran on edge datacenters. 
Tenants have different targets, and upon instantiation of the cellular network, they might also require the instantiation of a near-RT \gls{ric} to execute a catalog of xApps tailored to their needs.
This section describes \neutran in detail, starting from the O-RAN architecture used as supporting infrastructure, then describing the \neutran components and the procedures of its automated workflow.

\textbf{An O-RAN Primer}.
O-RAN is a disaggregated approach to deploy mobile cellular networks built upon cloud-native principles.
It introduces standardized interfaces that facilitate interoperability among  disaggregated network elements
(e.g., \gls{cu}/\gls{du}/\gls{ru}), and \glspl{ric} to oversee and fine-tune the functionalities of the network~\cite{oran-wg1-arch-spec}.
\glspl{ric} operate at different time scales to enable data-driven closed-control loops and network management through custom applications, called rApps (for the \textit{non-RT} \gls{ric}) and xApps (in the \textit{near-RT} \gls{ric}).\footnote{The extension of control and inference to \textit{real-time} time scales (i.e., below $10$\:ms) is possible via \textit{dApps}~\cite{doro2022dapps}. These are applications deployed directly on the \gls{ran} nodes, e.g., \glspl{cu}/\glspl{du}, and that can access sensitive data, e.g., I/Q samples, that cannot be streamed out of such nodes because of privacy concerns, among others.}
%
These applications receive live \glspl{kpm} from the \gls{ran}, and adapt its configuration to the dynamic channel conditions and traffic demand.
O-RAN also enables the deployment of virtualized services for the \ran in a pool of compute resources (the O-Cloud) managed by the \gls{smo} (Figure~\ref{fig:architecture}), a centralized component deployed in a cloud facility~\cite{polese2023understanding}. 
The \gls{smo} provides an abstract view of the network infrastructure and resources (e.g., compute, spectrum, coverage) obtained by using the O-RAN~O1 interface. It also triggers new service deployment and updates through the O2 interface, which connects the \gls{smo} to virtualization resources in the O-Cloud. The \gls{smo} hosts the O-RAN non-RT \ric and its rApps.

\textbf{\neutran components}.
\neutran consists of 
three main architectural components: (i)~an \gls{smo}; (ii)~edge datacenters, and (iii)~cell sites.
The \emph{\gls{smo}} in the \neutran architecture includes an instance of non-RT \gls{ric} with an rApp implementing the optimization engine (see Section~\ref{sec:formulation}).
%
Inputs to the engine include tenant requests and analytics from the \ran gathered by a monitoring rApp. 
Requests from the \neutran tenants are then matched into services to be deployed in the O-Cloud through the OpenShift Kubernetes \glspl{api}.
Other \neutran rApps automate service instantiation, and infrastructure monitoring for self-healing purposes. Last, a coverage rApp monitors historical and current coverage data to identify areas covered by each cell site (a key step in the optimization process, as described in Section~\ref{sec:formulation}).
%

\begin{figure}[ht]
\vspace{-.15cm}
    \centering
    \includegraphics[width=0.95\columnwidth]{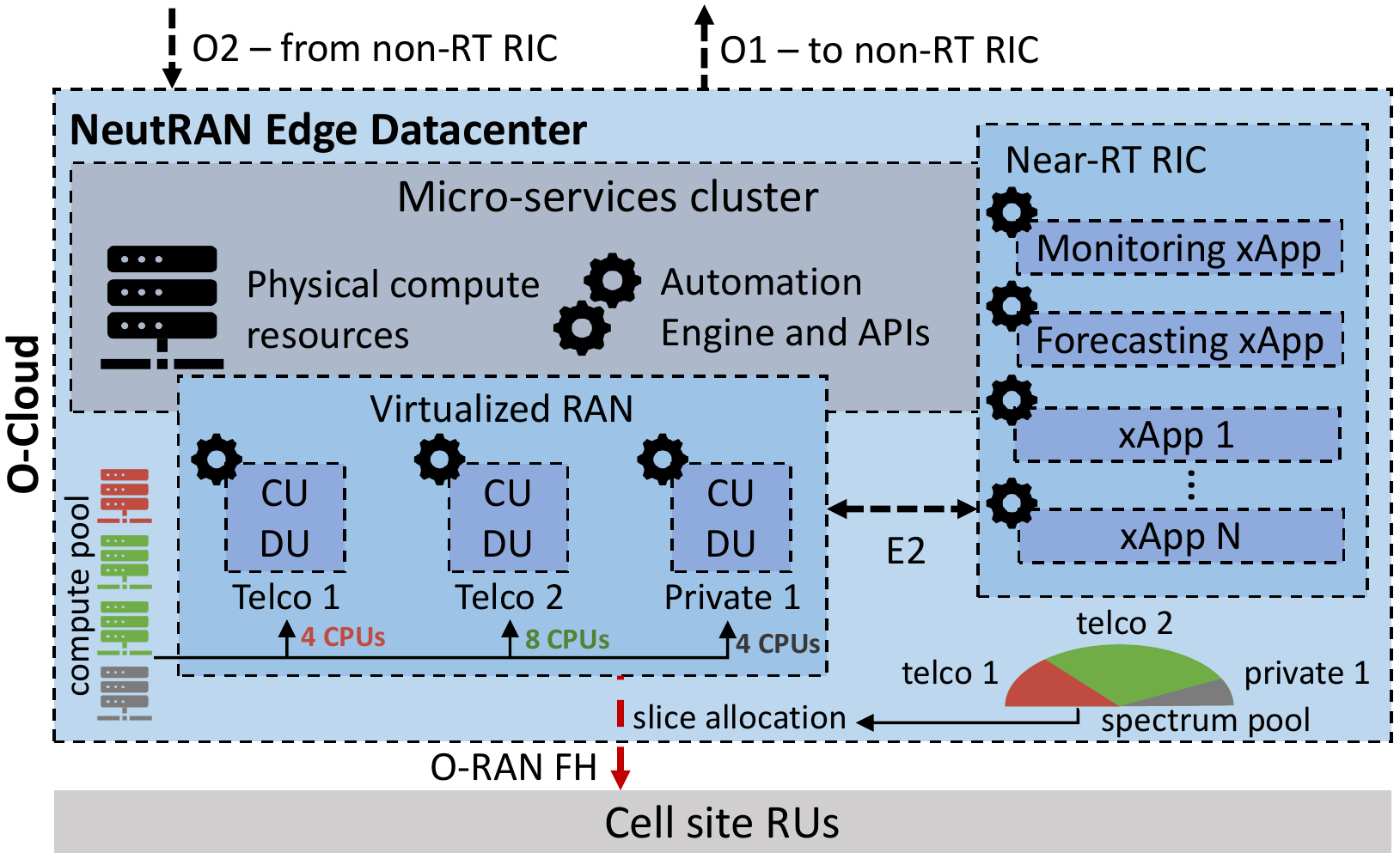}
    \setlength\belowcaptionskip{-.15cm}
    \caption{The \neutran edge datacenter, with a pool of shared compute and spectrum resources. The \neutran pipelines define how resources are sliced into \ran services and spectrum allocations for the tenants.}
    \label{fig:edge-datacenter}
\end{figure}

The \neutran \emph{edge datacenters} are illustrated in Figure~\ref{fig:edge-datacenter}. 
As part of the O-Cloud resource pool, they are implemented through OpenShift, the open-source enterprise-ready hybrid cloud platform-as-a-service framework by Red Hat~\cite{openshift}. 
OpenShift leverages containerized virtualization technologies managed by Kubernetes~\cite{kubernetes} to instantiate applications and workloads in the form of containers (or \textit{pods}) on top of white-box compute machines.
This framework also offers primitives and \glspl{api} to instantiate and manage the life cycle of custom workloads on top of the managed infrastructure.
Typical workloads include the virtualized core network and \ran, with the \glspl{cu} and \glspl{du} for different tenants, the near-RT \ric, connected to the virtualized \ran through the E2 interface, and the xApps running on the near-RT \ric. \neutran xApps aid and augment the operations of the optimization engine by providing additional monitoring and forecasting of user demand and resource utilization. 
The combination of custom OpenShift pipelines and of the directives from the optimization engine rApp enable efficient slicing of the edge datacenter resources, e.g., compute for \neutran services and spectrum to be used by the \glspl{cu} and \glspl{du} (Figure~\ref{fig:edge-datacenter}).
As we will discuss in Section~\ref{sec:prototype}, automated pipelines in the \neutran edge datacenters can start customized network services and workloads in less than $10$\:s.
The edge datacenters also expose Flask REST \glspl{api} to let the monitoring rApp query the current resource availability (e.g., compute, spectrum) in the OpenShift clusters.

\neutran datacenters are connected to multiple \emph{cell sites} in specific geographic areas via high-speed fiber connections, e.g., the O-RAN \gls{fh} interface. Cell sites host \glspl{ru} and antennas to provide \ran access over multiple frequency bands.

\textbf{The \neutran Automated Workflow}.
The \neutran end-to-end automated workflow is structured in six steps, shown in Figure~\ref{fig:architecture}. 
In \emph{Step~1}, tenants submit their \emph{requests} to \neutran through intents that describe the services required by the tenant (e.g., to cover a specific geographic area in a certain period of time), needed resources (e.g., spectrum), and fault-recovery policies (e.g., whether to re-instantiate services in case of failure).
These requests, together with spectrum and infrastructure availability, are the input to the rApp in the \gls{smo} that implements the optimization engine (\emph{Step~2}), whose outputs are allocation policies sent to the edge datacenters through the O2 interface in \emph{Step~3}. These policies specify the spectrum (carrier frequency and bandwidth) allocated to each tenant, together with the cell sites and compute resources in the edge datacenters.
In \emph{Step~4}, the \neutran edge datacenter shown in Figure~\ref{fig:edge-datacenter} uses automated pipelines programmed through OpenShift to dispatch services such as the \glspl{cu}/\glspl{du}, core network, and near-RT \glspl{ric} required by the tenants, together with the xApps in the
tenant's catalog. Upon instantiation, \ran services automatically connect to the core network and near-RT \ric running in the edge datacenter, report run-time \glspl{kpm} to the 
\ric through the O-RAN E2 interface, and expose functionalities the xApps can subscribe to.
Note that, based on the optimal allocation policy, multiple tenants can share the same base station. In this case, spectrum is shared by means of \gls{ran} slicing, where each tenant is assigned a different slice of the network (e.g., a subset of the available bandwidth). \neutran allocations are not elastic, i.e., between consecutive iterations of Step 3, which happen with a non-real-time periodicity, there is a rigid allocation of spectrum resources for each tenant. However, it is possible to combine \neutran with a slicing xApp to dynamically change the allocation for tenants in the same cell site on a near-real-time loop.
%
By \emph{Step 5}, the services required by the tenants on shared \ran and spectrum are fully provisioned.
After the instantiation of these micro-services, in \emph{Step 6} the edge datacenters and \gls{smo} run monitoring xApps and rApps to perform health checks on the deployed micro-services and resources, and to recover them from potential failures, thus effectively making the \gls{ran} self-healing.

%
%

%
%

%
%
%

\section{The \neutran Optimization Engine rApp}
\label{sec:formulation}


This section describes the \neutran optimization engine. 
The engine is implemented as an rApp (Figure~\ref{fig:architecture}) that models the \emph{neutral host problem} with data analytics from the \gls{smo} and \ran (Section~\ref{sec:model}).
Within the rApp, \neutran defines the problem, its constraints (Section~\ref{sec:definition}), and solves it (Section~\ref{sec:prob-formulation}) via reformulation-linearization techniques, discussed in Section~\ref{sec:solving}, where we also analyze in detail the problem complexity, and outline strategies for its reduction.
%
%

\subsection{System Model}
\label{sec:model}

We consider the \ran infrastructure in Figure~\ref{fig:architecture} with a set $\bss=\{1,2,...,B\}$ of $B$ cell sites and \neutran edge datacenters. This infrastructure is offered to a set $\tenants$ of $T$ tenants according to the neutral host business model: each cell site can be leased to a tenant for a set amount of time to offer network access to their users.
%
%
The deployment area is partitioned into a set $\areas=1,2,...,A$ of $A$ areas.
Each cell site covers one or more areas. 
%
For each area~$j\in\areas$ and cell site $b\in\bss$, the indicator variable~$c_{b,j}\in\{0,1\}$ is $c_{b,j}=1$ if $b$ provides coverage to $j$, and $c_{b,j}=0$ otherwise. 
%
We do not make assumptions on how indicator variables $c_{b,j}$ are computed, as different deployments may determine coverage with different policies. However, these can be used to enforce minimum \gls{qos} requirements to ensure that a base station $b$ is considered to be ``covering" a certain area $i$ only if it can deliver a satisfactory performance level to users deployed therein. For example, a realistic approach could set $c_{b,j}=1$ if and only if a metric~$\gamma_{b,j}$ (e.g., \gls{sinr}, throughput or channel quality) at cell site~$b$ for any user in area~$j$ exceeds a minimum tolerable value~$\gamma^{\mathrm{min}}$, where both~$\gamma_{b,j}$ and~$\gamma^{\mathrm{min}}$ can be obtained via historical data from the monitoring and coverage rApps (Section~\ref{sec:overview}). Let $\areas_b\subseteq\areas$ be the set of areas covered by $b\in\bss$.

\neutran uses a set~$\bands$ of~$W$ 5G frequency bands.
%
Let~$\freqs_\omega$ be the set of frequencies in band $\omega\in\bands$. 5G systems rely upon an \gls{ofdm} frame structure, which partitions frequencies into subcarriers. These are then organized into blocks of~12 to form the so-called \prbs, which are the minimum units that can be scheduled in frequency. As a consequence, we discretize the set~$\freqs_\omega$, which is the set of \prbs in band~$\omega$. Therefore, 
$\bigcup_{\omega=1}^W \freqs_{\omega} = \freqs$.

Each cell site can operate across multiple bands.
Thus, we introduce a variable~$\beta_{b,f}\in\{0,1\}$ such that, 
for any band~$\omega$ and cell site~$b$, $\beta_{b,f} = 1$ for all~$f\in\freqs_\omega$ if~$b$ can operate on band~$\omega$. We also introduce the indicator~$\xi_b\in\{0,1\}$, such that~$\xi_b=1$ if cell site~$b$ can transmit on a single band (among the bands supported by the cell site) at any given time.
%

We consider the case where each tenant in $\tenants$ can submit requests to provide wireless services 
at different locations. Specifically, each tenant $t$ in $\tenants$ generates a set~$\reqs_t$ of requests that are then collected into a set~$\reqs = \bigcup_{t\in\tenants} \reqs_t$ with a total of $I$ requests. As shown in Figure~\ref{fig:architecture}, each request $i\in\reqs$ specifies the area $j\in\areas$ where the service is needed, the required amount of resources $\delta_i$, and the level of fault resiliency. Without loss of generality, we consider a one-to-one mapping between a request $i$ and its associated area $j$.
%
$\delta_i$ represents the amount of \prbs required to accommodate the request. Its value might depend on a variety of factors such as number of users, their type (e.g., best-effort, premium), the type of traffic they generate (e.g., video streaming, browsing) as well as any \glspl{sla} in place between tenants and their customers. We consider two cases. In the first case, $\delta_i$ is submitted directly together with request $i$ by the tenants (either by a human or by an intelligent component, e.g., a request forecasting rApp) hosted in the \gls{smo} or in the non-RT \gls{ric}. This demand might depend on market and business strategies, and its value is usually kept undisclosed out of privacy concerns. As such, we do not make any assumptions on how tenants compute the value of~$\delta_i$, and design \neutran to let the tenants keep the models used to compute $\delta_i$ undisclosed, having only to specify its value.
In the second case, we follow an approach similar to that in \cite{sciancalepore2017traffic} where operators specify the type of traffic, slice and corresponding \gls{sla} levels only, and \neutran can predict and compute the amount of resources $\delta_i$ that are required to satisfy such request.

Without loss of generality, we consider the case where requests in $\reqs$ are collected by \neutran following a timeslot-based approach. Specifically, tenants are allowed to submit their request at any given time, but \neutran will process such requests simultaneously every $\Delta$ seconds. As a consequence, the outstanding requests are stored into a buffer (i.e., the outstanding request set $\reqs$). Every $\Delta$ seconds, \neutran computes an optimal solution and removes the satisfied requests from the outstanding set $\reqs$. Tenants whose requests have not been selected can decide to either keep their request in the buffer or remove them. As we will show in Section \ref{sec:prototype}, it takes about 9 seconds to activate a base station, while in Section \ref{sec:numerical} we will show how \neutran can compute optimal solutions within a few couple of seconds. Therefore, the duration of $\Delta$ in most cases is lower-bounded by the time to activate \gls{ran} components rather than the time to compute an optimal solution.

As cell sites offer limited coverage, we define two sets $\bss_{i}$ and $\bss_{-i}$ to represent the cell sites that offer coverage to the area~$j$ specified by request $i$ and those that do not, respectively.
%
%
These sets are defined as $\bss_{i} = \{ b \in \bss : r_{i,b} = 1 \} \subseteq \bss$ and $\bss_{-i} = \bss\setminus{\bss_{i}}$, where $r_{i,b}\in\{0,1\}$ is a variable used to determine whether or not cell site $b$ is a suitable candidate to accommodate request $i$. Specifically, for any request $i$ and its required area $j$, we have that $r_{i,b}=1$ if and only if $j\in\areas_b$.
Finally, we also introduce a parameter~$w_i$ that can be used to model the value of request~$i$. 
This adds flexibility to \neutran, as it can be used by tenants to declare the monetary value of the requests, and by \neutran to prioritize profits over infrastructure~utilization.

\subsection{Problem Definition} \label{sec:definition}

\neutran is designed to enable neutral host applications for Open \gls{ran} cellular systems offering at the same time an automated and optimized platform to (i) instantiate disaggregated 5G \glspl{gnb}; (ii) allocate spectrum on-demand; (iii) avoid interference; and (iv) satisfy tenant requests. 

%
Let $\mf{y} = (y_{i,b})_{i\in\reqs, b\in\bss}$, and $\mf{x} = (x_{i,b,f})_{i\in\reqs, b\in\bss, f\in\freqs}$ be our optimization variables. 
%
%
%
In our formulation, 
variable~$y_{i,b}\in\{0,1\}$ is set to~$1$ if request $i$ is assigned to cell site~$b$; $y_{i,b}=0$ otherwise. 
Variable $x_{i,b,f}\in\{0,1\}$ indicates which \prbs on cell site~$b$ have been allocated to request~$i$, namely, \mbox{$x_{i,b,f}=1$} indicates that \prb~$f$ in cell site~$b$ is assigned to request~$i$.
%
The constraints and objective of the optimization are as follows.

\subsubsection{Avoid conflicts and spectrum over-provisioning}
\label{sec:model:definition:interference}

First, each request~$i$ can be allocated to one cell site~$b$ only to ensure that the infrastructure owner does not incur additional costs by instantiating many cell sites to satisfy the same request. 
For all~$i\in\reqs$, this is formulated as follows:
\begin{equation}
    \sum_{b\in\bss} y_{i,b} \leq 1
    \label{eq:con:one_bs}
\end{equation}
Furthermore, we must also ensure that \neutran (i)~does not allocate more \prbs than available; (ii)~avoids conflicts by not allocating the same \prb to multiple requests (e.g., from different tenants); and (iii)~mitigates interference by making sure neighboring cells do not operate over the same spectrum. 

For each~$b$ and~$b'\in\bss$, let the interference indicator $I_{b,b'}\in\{0,1\}$ be such that $I_{b,b'} = 1$ if~$b$ and~$b'$ have overlapping coverage regions and interfere with each other if using the same spectrum.
For all~$f\in\freqs$, and $b,b' \in \bss$ with $I_{b,b'} = 1$, the following two inequalities model constraints~(i)+(ii), and constraint~(iii), respectively.
\begin{align}
    \sum_{i\in\reqs} x_{i,b,f}  & \leq \beta_{b,f}
    \label{eq:con:overprovisioning} \\ \sum_{i\in\reqs} (x_{i,b,f} \beta_{b,f} + x_{i,b',f} \beta_{b',f})  & \leq 1
    \label{eq:con:zero_intf}
\end{align}
Note that the left-hand side of~\eqref{eq:con:overprovisioning} ensures that only supported bands can be allocated, and that the total number of \prbs allocated to each cell site does not exceed the total number of available \prbs. 
This is because~$F_b = \sum_{f\in\freqs} \beta_{b,f}$ indicates the number of \prbs available at~$b$ across all supported bands. Therefore, since~$\beta_{b,f} \leq 1$, \eqref{eq:con:overprovisioning} enforces the allocation of no more than the available $F_b$ \prbs. 
%

\subsubsection{Satisfy locality and spectrum demand}

To satisfy request $i$, \neutran can select any cell site $b$ that covers area $j$ requested by $i$, i.e., $r_{i,b}\!=\!1$, and must satisfy the spectrum demand $\delta_i$ by allocating enough \prbs to the request. These constraints can be defined jointly via~\eqref{eq:con:demand} for all $i\in\reqs$ and $b\in\bss$.
\begin{equation}
    \sum_{f\in\freqs} x_{i,b,f} \beta_{b,f} r_{i,b} = \delta_{i} y_{i,b}
    \label{eq:con:demand}
\end{equation}
%
%
For all~$i\in\reqs$ and~$b\in\bss$, the following set of constraints~\eqref{eq:con:force_zero} forces to zero all variables that would result in an unfeasible solution where a request~$i$ is allocated to a cell site~$b$ that does not offer coverage to the required area, i.e., $b\in\bss_{-i}$. 
\begin{equation}
    \sum_{b'\in\bss_{-i}} y_{i,b'} + \sum_{b'\in\bss_{-i}} \sum_{f\in\freqs} x_{i,b',f} = 0
    \label{eq:con:force_zero}
\end{equation}
%

\subsubsection{Enforce contiguous allocation}



Allocating contiguous \prbs to requests makes it possible to implement the sharing mechanism through \ran slicing, and decreases the complexity of transceiver architectures, 
e.g., it eliminates the need for carrier aggregation~\cite{foukas2017orion}.
For all $i\in\reqs$ and $b\in\bss$, the contiguous allocation of \prbs can be enforced via the  following constraint:
\begin{equation}
    \sum_{f = 1}^{F-1} x_{i,b,f} \cdot x_{i,b,f+1} \alpha_{f,f+1} = (\delta_{i} - 1) y_{i,b}
    \label{eq:con:contiguous}
\end{equation}
\noindent
where $\alpha_{f,f'}\in\{0,1\}$ is such that $\alpha_{f,f}=0$ and $\alpha_{f,f'}=1$ if $f$ and $f'$ belong to the same spectrum band $\omega$, i.e., $(f,f') \in \freqs_\omega \times \freqs_\omega$, and they are consecutive, i.e., $f'=f\pm1$. In this way, together with \eqref{eq:con:demand}, \eqref{eq:con:contiguous} ensures that $y_{i,b}=1$ if and only if we allocate exactly $\delta_i$ contiguous \prbs.  

\subsubsection{Support single band cell sites}

%
Smaller cell sites (e.g., micro or picocells) may support multiple spectrum bands, but can transmit over a single spectrum band only at any given time.
This must be enforced through equality~\eqref{eq:con:one_band_only} which, for all $b\in\bss$ and $\omega \in \bands$ such that $\xi_b = 1$ (i.e., those~$b$ that only support single band operations), ensures that only one band is allocated to requests at any given time.
\begin{equation}
    \sum_{f\in\freqs_{\omega}}\sum_{i\in\reqs} x_{i,b,f} \cdot \sum_{f'\in\freqs\setminus{\freqs_{\omega}}}\sum_{i\in\reqs} x_{i,b,f'} = 0
    \label{eq:con:one_band_only}
\end{equation}

\subsection{Problem Formulation}
\label{sec:prob-formulation}

The neutral host optimization problem is as follows:
\begin{align}
    \underset{{\mf{x},\mf{y}}}{\maximize} \hspace{0.5cm} & \sum_{i\in\reqs} \sum_{b\in\bss} y_{i,b} w_i \label{eq:optimization}\\
    \mathrm{subject~to} \hspace{0.5cm} & \mbox{Constraints}~ \eqref{eq:con:one_bs}, 
    \eqref{eq:con:overprovisioning},
    \eqref{eq:con:zero_intf},
    \eqref{eq:con:demand}, 
    \eqref{eq:con:force_zero},
    \eqref{eq:con:contiguous},
    \eqref{eq:con:one_band_only}
    \nonumber \\
    & x_{i,b,f} \in \{0,1\}, \hspace{0.2cm} \forall i\in\reqs,~b\in\bss, f\in\freqs \nonumber \\
    & y_{i,b} \in \{0,1\}, \hspace{0.5cm} \forall i\in\reqs,~b\in\bss \nonumber
\end{align}
The objective of \eqref{eq:optimization} is to accommodate as many requests as possible such that their cumulative value is maximized while satisfying the set of constraints discussed in Section \ref{sec:definition}. 
%
We notice that a specific instance of \eqref{eq:optimization} is that where $w_i=1$ for all $i\in\reqs$. In this case, the formulation would allow infrastructure owners to maximize the number of admitted requests rather than their cumulative value.

\subsection{Complexity Analysis and Mitigation} 
\label{sec:solving}

The neutral host problem formalized above is a \emph{binary Quadratically Constrained Quadratic Program} (QCQP), as the optimization variables~$\mf{x}$ and~$\mf{y}$ are~0-1 variables and constraints~\eqref{eq:con:contiguous} and~\eqref{eq:con:one_band_only} are quadratic.
%
%

In general, even non-binary QCQPs are NP-hard~\cite{huang2016consensus}, and their binary 0-1 version results in a non-convex optimization problem that inherits the NP-hardness of binary quadratic problems~\cite{androulakis2008quadratic}, which make binary QCQPs---such as problem \eqref{eq:optimization}---NP-hard.
Despite the exponential complexity of such problems, Semidefinite Programming Relaxations (SPR)~\cite{boyd1997semidefinite} and Reformulation-Linearization Techniques (RLT)~\cite{sherali2013reformulation} have been shown to be effective tools for solving them optimally~\cite{anstreicher2009compare}.
%
Both SPR and RLT transform the non-linearity from the product of any two variables $z_n$ and $z_m$ 
(such as those in \eqref{eq:con:contiguous} and \eqref{eq:con:one_band_only}) 
into an auxiliary variable $Z_{n,m}=z_n z_m$. However, the two approaches differ in that each technique adds a different set of constraints to the problem. We refer the reader to \cite{anstreicher2009compare} for a detailed comparison of the two approaches. 
%
In this paper, we solve problem~\eqref{eq:optimization} in Gurobi and use RLT, which has been shown to deliver faster computing time and higher accuracy than SPR in large-scale configurations~\cite{basu2017large}.



The primary source of complexity of problem~\eqref{eq:optimization} stems from the number of variables $N$ of the problem, i.e., $N=N_x + N_y = IBF + IB \in \mc{O}(IBF)$. 
Note that, while the number of requests $I$ and the number $B$ of cell sites might be arbitrarily large, the total number $F$ of \prbs is upper-bounded by $F^{\mathrm{MAX}} = 275\cdot W$, if all the $W$ spectrum bands in $\bands$ support the maximum number of \prbs allowed by 5G NR~\cite{3gpp.38.211}.
For instance, in a scenario with $B=5$ cell sites, $W=10$ bands, each with~275 \prbs, and $I=10$ requests, there are more than $130,\!000$ optimization variables and more than $600,\!000$ constraints from~\eqref{eq:con:zero_intf} alone.

Therefore, we have designed the following complexity reduction and relaxation techniques:

\noindent$\bullet$~\textit{Variable Reduction (VR):} any variable can be eliminated that: (i)~always results in unfeasible solutions, or (ii)~is always equal to zero due to the structure of the problem. For example, for any request $i$, we have that allocating $i$ to any cell site $b\in\bss_{-i}$ is unfeasible. Similarly, if cell site $b\in\bss_{i}$ but a specific band $\omega\in\bands$ is not supported by the cell site, then any $x_{i,b,f}$ with $f\in\freqs_\omega$ will always be equal to zero due to \eqref{eq:con:overprovisioning}. For this reason, all of these variables can be removed from the search space, thus reducing the time necessary to compute an optimal solution.

\noindent$\bullet$~\textit{\prb grouping (PG):} \prbs can be bundled together into groups of minimum size~$K$ such that the problem is cast into a space with $\tilde{F} = F/K$ \prbs. A preliminary grouping is naturally occurring in any~5G NR systems as \prbs are grouped into \glspl{rbg} with varying numbers of \prbs, depending on the specific numerology.
%
We can further extend this concept in at least two ways. The first one assumes that as tenants submit their requests to the infrastructure owner that groups \prbs by computing the greatest common divisor (GCD) among all demands $\delta_i$  such that $K=\mathrm{GCD}(\reqs)$.
Another approach instead would force tenants to submit requests with a demand whose value is a multiple of a fixed block size~$K$, i.e., $\delta_i = n_i  K$, with~$n_i$ an integer for all $i\in\reqs$.
While the first approach leaves more room for customizing requests to the tenants (which can potentially generate requests for a single \prb), it is prone to achieve poor complexity reduction as the GCD might be small. On the contrary, the second approach is less flexible for the tenants but results in lower computational complexity as the infrastructure owner can select a large~$K$.


\subsubsection{On the optimality gap}

The impact of VR and PG on the optimality of the solution is worth discussing. First, VR eliminates only \textit{inactive variables}, i.e., those variables of the problem that would be equal to 0 due to structural properties. For example, VR would eliminate those variables that would force allocating $i$ to any cell site $b\in\bss_{-i}$. In this case, it is trivial to prove that VR has no impact on the optimality of the  solution as the removed variables do not contribute to the objective function, and do not affect the solution, which remains optimal and with an optimality gap of zero. PG, instead, affects the way requests are being generated and processed by \neutran. When using PG \neutran forces tenants to either request a number of \prbs that is a multiple of a fixed block size~$K$, or groups resources using their GCD $K$ in an effort to convert requested resources into blocks of requested resources. In both cases, the solution computed by \neutran is still optimal with respect to the requests being submitted, as the allocation always satisfies the requests and only changes the way they are being allocated (i.e., individually or as a group).

\section{The \neutran Prototype}
\label{sec:prototype}

\begin{figure}[t]
    \setlength\belowcaptionskip{-.3cm}
    \centering
    \includegraphics[width=\columnwidth]{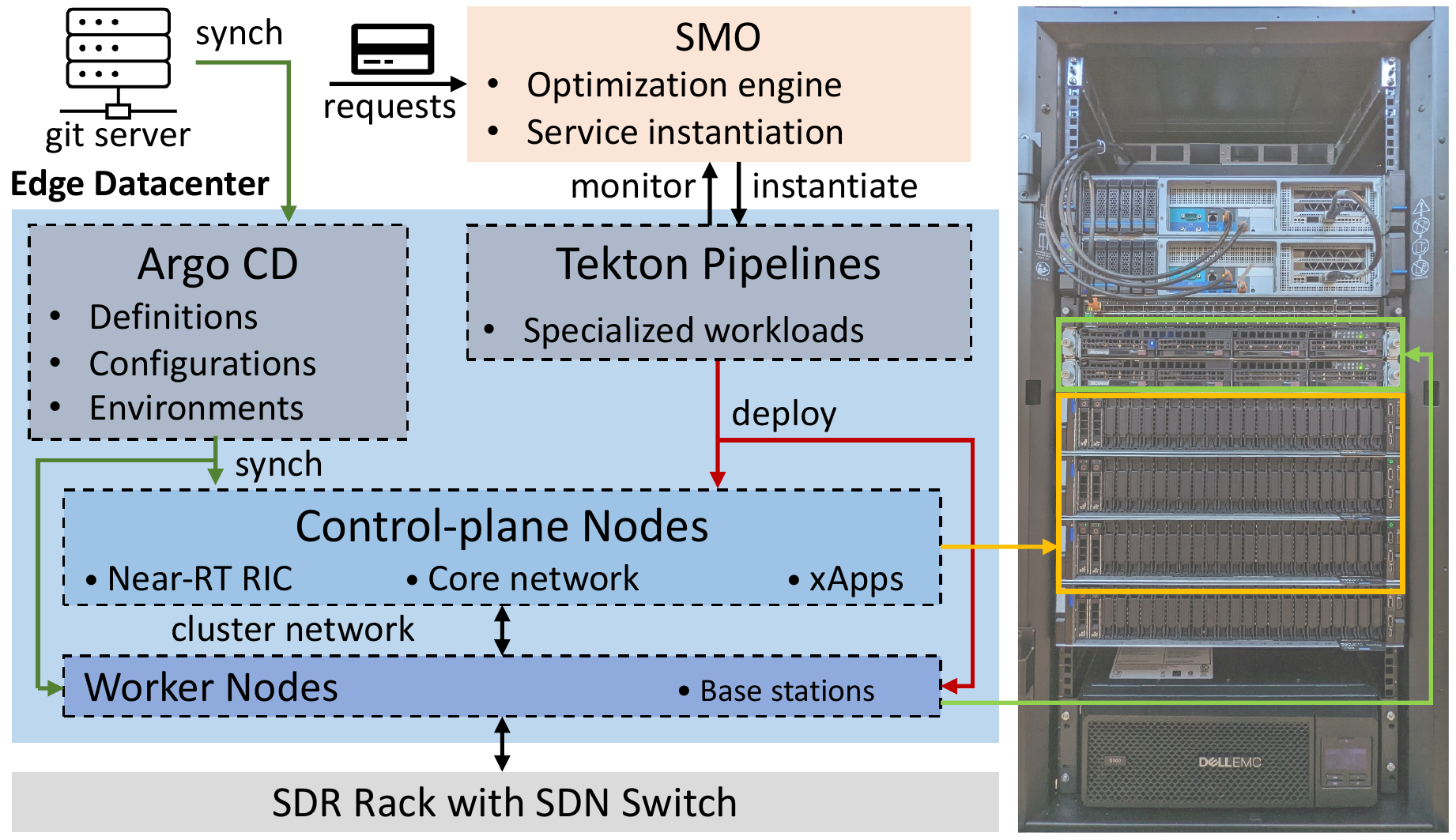}
    \caption{The OpenShift-based \neutran prototype.}
    \label{fig:prototype}
\end{figure}

The \neutran prototype (Figure~\ref{fig:prototype}) implements the framework described in Sections~\ref{sec:overview} and~\ref{sec:formulation}. It features unique software and hardware components that enable automated pipelines for end-to-end, optimization-driven spectrum and \ran sharing.

\textbf{Software---automation.} 
The overall software infrastructure involves more than~330 components executing as either micro-services deployed as OpenShift pods (i.e., containerized applications) or rApps in the \gls{smo}. To develop our \neutran prototype and enable the seamless transition from tenant requests (specified via a graphical control interface) to deployment of fully operational cellular networks, we have implemented a set of custom automation pipelines on top of the edge datacenter OpenShift infrastructure using open-source, cloud-native continuous integration and delivery frameworks. These make it possible to apply the output of the optimization engine rApp (in the \gls{smo}) to \emph{generic application templates} (see Listing~\ref{lst:deployment-template} for an example) and translate them into \emph{custom services} that are then automatically deployed on the cluster. 

Notably, the \gls{smo} uses an O2-like interface to control the OpenShift \glspl{api} and the various services deployed on the edge datacenter. 
The optimization engine rApp computes the optimal allocation of \gls{ran} services by solving the neutral host problem described in Section~\ref{sec:formulation}, and automatically instantiates the resulting services (e.g., the \ran applications) by adapting a set of generic templates to the specific requested services at run time.
Templates are deployed on the OpenShift cluster through the Argo \gls{cd} framework, which supports declarative application definitions, configurations, and environments synchronized from a version-controlled source (e.g., a \texttt{git} server)~\cite{argocd_website}.
Starting from these templates, the actual workloads and services resulting from the \neutran optimization are instantiated as pods from an internal Docker image registry through Tekton pipelines~\cite{tekton_website}.

\begin{lstlisting}[float=t,floatplacement=b,language=yaml,style=mystyle-yaml,
caption={Base station deployment template to be specialized at run time.},
label={lst:deployment-template}]
apiVersion: template.openshift.io/v1
kind: Template
metadata: # template name (e.g., neutran) and annotations
parameters: # template parameters (e.g., frequencies; core network, RIC, and USRP IP; slice allocations)
objects:
  - kind: Deployment
    apiVersion: apps/v1
    metadata: # deployment name (e.g., neutran-cell-1), namespace (e.g., neutran), and labels
    spec:
      template:
        metadata: # template labels, and annotations
        spec:
          nodeSelector: # to select low-latency nodes
          containers:
            - name: # pod name (e.g., neutran-cell-1)
              image: # Docker image (e.g., neutran-cell)
              command: # pod-entrypoint (e.g., /run.sh)
              env: # parameteres of line 4 passed as environment variables
              ports: # exposed ports and protocols
              resources: # pod compute resources
  - kind: Service # exposed services (e.g., Flask APIs)
  - kind: Route # routes to reach the exposed services
\end{lstlisting}

After their instantiation, applications and services are actively monitored by \neutran, which can tune their configuration at run time based on subsequent optimization results, and re-instantiate them if necessary (e.g., in case of conflicts between services, or failure of a certain service).
In this way, \neutran is resilient to failures, and self-adapts to heterogeneous network deployments and operator requests.

\textbf{Software---edge services.} We deployed the functionalities that support OpenShift on dedicated control-plane nodes (e.g., cluster monitoring services, operators, certificate managers, DNS, etc.), which also host additional edge micro-services. 
%
%
These include (i)~an E-release O-RAN near-RT \gls{ric} provided by the \gls{osc}, with an E2 termination to the \ran for data collection and performance reporting; (ii)~data-driven xApps running on the near-RT \gls{ric}; and (iii)~a core network implemented through Open5GS~\cite{open5gs_website}.
We configured additional compute resources (worker nodes) to only execute low-latency applications, i.e., the base stations, thus providing performance guarantees.
The \gls{ran} is implemented through the \scope software-defined cellular stack, part of the publicly available \openrangym framework~\cite{bonati2022openrangym-pawr}. \scope extends srsRAN with network slicing capabilities (leveraged to implement spectrum sharing among different tenants), and the O-RAN-compliant E2 termination to communicate with the near-RT \ric~\cite{bonati2021scope, srsran_website}. Every application is containerized and exposes Flask REST \glspl{api} for monitoring and re-configuration.

\textbf{Hardware.} We deployed the three main infrastructure components of \neutran as follows: (i) the \gls{smo}, on an Intel NUC (15~CPU cores, 64\:GB RAM); (ii) the edge datacenter, on a bare-metal cluster managed by OpenShift; and (iii) four cell sites, on USRPs X310 part of an infrastructure with \glspl{sdr}, antenna locations, and computational facilities.
The cluster features three control-plane nodes (Dell PowerEdge R740, 32 CPU cores and 192\:GB RAM) and two worker nodes (Microway EPYC, 32 CPU cores and 256\:GB RAM).
To ensure low latency and high performance, each node embeds a $100$\:Gbps Ethernet card from NVIDIA Mellanox. Workers connect to the \glspl{sdr} via a Dell 4048T-ON \gls{sdn} switch.

An extensive experimental evaluation of \neutran and its procedure is presented in Section \ref{sec:exp-results}, where we demonstrate that \neutran can effectively deploy a fully virtualized network in less than 10 seconds while improving a variety of \glspl{kpm}.
We notice that \neutran can deploy an end-to-end cellular network on white-box infrastructure in less than $10$\:s, demonstrating its feasibility and effectiveness.

\section{\neutran Performance Evaluation} 
\label{sec:evaluation}

This section presents the \neutran performance evaluation, including numerical results for scalability and resource utilization, and experimental results about the gain that sharing introduces for multiple \glspl{kpm}.

\subsection{Scalability, Complexity, and Effectiveness}
\label{sec:numerical}


We run large-scale simulations to (i)~evaluate the computational complexity and scalability of \neutran, and (ii)~assess and characterize its performance to determine its applicability to real-world large-scale deployments. 
%
%

The following results are generated through a custom MATLAB simulator that uses Gurobi to solve problem \eqref{eq:optimization} on a compute node with $32$\:GB of RAM and a 12-core Intel Core i7-9750H CPU at $2.60$\:GHz. We consider a deployment with 5G NR cell sites uniformly deployed on a grid with $A=21\times11=231$ areas.
The \ran uses numerology~4, i.e., $240$\:kHz subcarrier spacing and 138 \glspl{prb} per band~\cite{3gpp.38.211}. 
The coverage indicators $c_{b,j}$ are configured such that $c_{b,j}=1$ if area $j\in\areas$ is distant at most $3\sqrt{2}/2$ from the cell site, where the distance is normalized with respect to the width of the area. We consider the case where $w_i=1$ (i.e., problem \eqref{eq:optimization} aims at maximizing the number of admitted requests). All results
are averaged over~100 independent simulation runs.

\textbf{Complexity Analysis.} 
%
%
Here, we leverage and profile the reduction techniques VR and PG presented in Section~\ref{sec:solving}. They identify the optimal solution for problem~\eqref{eq:optimization} with a reduced complexity, avoiding computations of $60$\:s or more even in scenarios with few cells or bands.


\ifexttikz
    \tikzsetnextfilename{complexity}
\fi
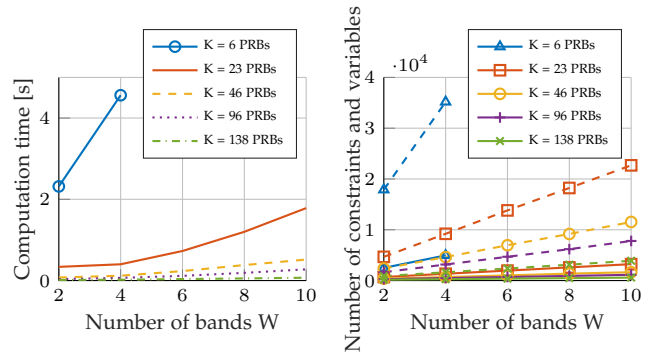
\begin{figure}[b]
    \centering
    \setlength\fwidth{0.9\columnwidth}
    \setlength\fheight{.32\columnwidth}
%
%
\definecolor{mycolor1}{rgb}{0.00000,0.44700,0.74100}%
\definecolor{mycolor2}{rgb}{0.85000,0.32500,0.09800}%
\definecolor{mycolor3}{rgb}{0.92900,0.69400,0.12500}%
\definecolor{mycolor4}{rgb}{0.49400,0.18400,0.55600}%
\definecolor{mycolor5}{rgb}{0.46600,0.67400,0.18800}%
\definecolor{mycolor6}{rgb}{0.00000,0.44706,0.74118}%
\definecolor{mycolor7}{rgb}{0.85098,0.32549,0.09804}%
\definecolor{mycolor8}{rgb}{0.92941,0.69412,0.12549}%
\definecolor{mycolor9}{rgb}{0.49412,0.18431,0.55686}%
\definecolor{mycolor10}{rgb}{0.46667,0.67451,0.18824}%
\begin{tikzpicture}
\pgfplotsset{every tick label/.append style={font=\scriptsize}}

\begin{axis}[%
width=0.411\fwidth,
height=0.95\fheight,
at={(0\fwidth,0\fheight)},
scale only axis,
xmin=2,
xmax=10,
xlabel style={font=\footnotesize\color{white!15!black}},
xlabel={Number of bands W},
ymin=0,
ymax=5,
extra y ticks={5},
extra y tick labels={},
extra y tick style={tick style={draw=none}},
ylabel style={font=\footnotesize\color{white!15!black}},
ylabel={Computation time [s]},
axis background/.style={fill=white},
axis x line*=bottom,
axis y line*=left,
xmajorgrids,
ymajorgrids,
legend style={at={(0.34,0.6)}, anchor=south west, legend cell align=left, align=left, draw=white!15!black, font=\tiny},
xlabel shift=-2pt,
ylabel shift=-5pt
]
\addplot [color=mycolor1, thick, mark size=2pt, mark=o, mark options={solid, mycolor1}]
  table[row sep=crcr]{%
2	2.31666075\\
4	4.5615425\\
};
\addlegendentry{K = 6 PRBs}

\addplot [color=mycolor2, thick]
  table[row sep=crcr]{%
2	0.33706844\\
4	0.40036236\\
6	0.72776006\\
8	1.19980618\\
10	1.7832667\\
};
\addlegendentry{K = 23 PRBs}

\addplot [color=mycolor3, dashed, thick]
  table[row sep=crcr]{%
2	0.0745487\\
4	0.1187008\\
6	0.23348384\\
8	0.3810522\\
10	0.51667682\\
};
\addlegendentry{K = 46 PRBs}

\addplot [color=mycolor4, dotted, thick]
  table[row sep=crcr]{%
2	0.03425326\\
4	0.06536018\\
6	0.11575146\\
8	0.1893736\\
10	0.27599304\\
};
\addlegendentry{K = 96 PRBs}

\addplot [color=mycolor5, dashdotted, thick]
  table[row sep=crcr]{%
2	0.01505312\\
4	0.02019006\\
6	0.03627246\\
8	0.05228894\\
10	0.06970448\\
};
\addlegendentry{K = 138 PRBs}

\end{axis}

\begin{axis}[%
width=0.411\fwidth,
height=0.95\fheight,
at={(0.54\fwidth,0\fheight)},
scale only axis,
xmin=2,
xmax=10,
xlabel style={font=\footnotesize\color{white!15!black}},
xlabel={Number of bands W},
ymin=0,
ymax=40000,
ylabel style={font=\footnotesize\color{white!15!black}},
ylabel={Number of constraints and variables},
axis background/.style={fill=white},
axis x line*=bottom,
axis y line*=left,
xmajorgrids,
ymajorgrids,
legend style={at={(0.34,0.6)}, anchor=south west, legend cell align=left, align=left, draw=white!15!black, font=\tiny},
xlabel shift=-2pt,
ylabel shift=-5pt
]
\addplot [color=mycolor1, thick, mark size=2pt, mark=triangle, mark options={solid, mycolor1}]
  table[row sep=crcr]{%
2	2491\\
4	4952.25\\
};
\addlegendentry{K = 6 PRBs}

\addplot [color=mycolor2, thick, mark size=2pt, mark=square, mark options={solid, mycolor2}]
  table[row sep=crcr]{%
2	687.7\\
4	1341.5\\
6	1957.3\\
8	2592.1\\
10	3230.56\\
};
\addlegendentry{K = 23 PRBs}

\addplot [color=mycolor3, thick, mark size=2pt, mark=o, mark options={solid, mycolor3}]
  table[row sep=crcr]{%
2	366.38\\
4	688.74\\
6	1010.04\\
8	1363\\
10	1639.9\\
};
\addlegendentry{K = 46 PRBs}

\addplot [color=mycolor4, thick, mark size=2pt, mark=+, mark options={solid, mycolor4}]
  table[row sep=crcr]{%
2	265.5\\
4	481.68\\
6	696.54\\
8	910.18\\
10	1132.74\\
};
\addlegendentry{K = 96 PRBs}

\addplot [color=mycolor5, thick, mark size=2pt, mark=x, mark options={solid, mycolor5}]
  table[row sep=crcr]{%
2	161.52\\
4	265.7\\
6	373.1\\
8	475.92\\
10	577.28\\
};
\addlegendentry{K = 138 PRBs}

\addplot [color=mycolor6, dashed, thick, mark size=2pt, mark=triangle, mark options={solid, mycolor6}, forget plot]
  table[row sep=crcr]{%
2	17882\\
4	35178.5\\
};
\addplot [color=mycolor7, dashed, thick, mark size=2pt, mark=square, mark options={solid, mycolor7}, forget plot]
  table[row sep=crcr]{%
2	4667.08\\
4	9207\\
6	13813\\
8	18225.64\\
10	22685.92\\
};
\addplot [color=mycolor8, dashed, thick, mark size=2pt, mark=o, mark options={solid, mycolor8}, forget plot]
  table[row sep=crcr]{%
2	2366.28\\
4	4627.88\\
6	6951.2\\
8	9173.2\\
10	11536.6\\
};
\addplot [color=mycolor9, dashed, thick, mark size=2pt, mark=+, mark options={solid, mycolor9}, forget plot]
  table[row sep=crcr]{%
2	1645.4\\
4	3158.08\\
6	4681.88\\
8	6179.56\\
10	7793.48\\
};
\addplot [color=mycolor10, dashed, thick, mark size=2pt, mark=x, mark options={solid, mycolor10}, forget plot]
  table[row sep=crcr]{%
2	887.84\\
4	1629.96\\
6	2372.52\\
8	3120.32\\
10	3898.56\\
};
\end{axis}
\end{tikzpicture}%
    \setlength\abovecaptionskip{0.05cm}
    \caption{Computational complexity analysis for different number of spectrum bands ($W$) and grouping coefficients ($K$) by using both VR and PG. Left: computation time in seconds; Right: number of variables (solid lines) and constraints (dashed lines). $I=20, B=50$.}
    \label{fig:complexity}
\end{figure}

Figure~\ref{fig:complexity} offers an analysis of the computational complexity of problem~\eqref{eq:optimization} aimed at showing the scalability of \neutran.
We consider scenarios with variable numbers~$W$ of spectrum bands per cell site and different values of~$K$.
%
%
%
%
The figure on the left shows that \neutran can compute optimal solutions in few hundreds of milliseconds even in the case of large~$W$ and grouping coefficients~$K\geq23$ (which corresponds to dividing the available~138 \prbs in~6
groups of size~23). 
We notice that grouping with higher values of $K$ is indeed effective, as it enables a 97\% reduction in the computation times of optimal solutions (i.e., from $4.56$\:s of $K\!\!=\!\!6$ to $0.12$\:s of $K\!\!=\!\!23$ for $W\!=\!4$).
%
This is due to the reduction in the number of variables (solid lines) and constraints (dashed lines) required to solve problem \eqref{eq:optimization}, which are shown on the right side of Figure~\ref{fig:complexity}.

\textbf{Resource Utilization and Allocation Effectiveness.}
%
Another important aspect to investigate is
how many requests are admitted by \neutran, as well as
how many cell sites and spectrum bands are activated for different configurations and deployments. To capture real-world deployment characteristics, the probability $P_{\mr{NS}}$ models those cases where a cell site $b$ can support a number of bands $W_b \leq W$.
In each
run and for each $b\in\bss$ and band $\omega\in\Omega$, we generate a random variable $z$ from a uniform distribution in $[0,1]$ and set $\beta_{b,f}=0$ for each~$f\in\omega$ if~$z < P_{\mr{NS}}$.
%
Additionally, the value of the single band indicator variable~$\xi_b$ is modeled with a probability $P_{\mr{SB}}$: for each $b\in\bss$, we randomly draw the variable $z$ again and set $\xi_b=1$ if $z < P_{\mr{SB}}$. In the following, we consider $W=5$.


\ifexttikz
    \tikzsetnextfilename{pns}
\fi
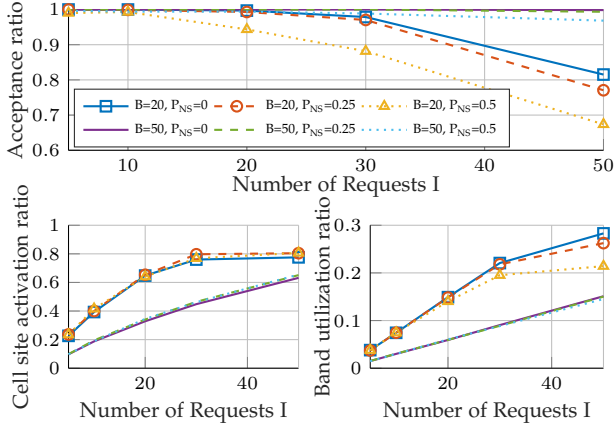
\begin{figure}[t]
    \centering
    \setlength\fwidth{0.8\columnwidth}
    \setlength\fheight{.55\columnwidth}
%
%
\definecolor{mycolor1}{rgb}{0.00000,0.44700,0.74100}%
\definecolor{mycolor2}{rgb}{0.85000,0.32500,0.09800}%
\definecolor{mycolor3}{rgb}{0.92900,0.69400,0.12500}%
\definecolor{mycolor4}{rgb}{0.49400,0.18400,0.55600}%
\definecolor{mycolor5}{rgb}{0.46600,0.67400,0.18800}%
\definecolor{mycolor6}{rgb}{0.30100,0.74500,0.93300}%
\begin{tikzpicture}
\pgfplotsset{every tick label/.append style={font=\scriptsize}}

\begin{axis}[%
width=\fwidth,
height=0.383\fheight,
at={(0\fwidth,0.543\fheight)},
scale only axis,
xmin=5,
xmax=50,
xlabel style={font=\footnotesize\color{white!15!black}},
xlabel={Number of Requests I},
ymin=0.6,
ymax=1,
ylabel style={font=\footnotesize\color{white!15!black}},
ylabel={Acceptance ratio},
axis background/.style={fill=white},
axis x line*=bottom,
axis y line*=left,
xmajorgrids,
ymajorgrids,
legend style={at={(0.01, 0.02)}, anchor=south west, legend cell align=left, align=left, draw=white!15!black, font=\tiny},
legend columns=3,
xlabel shift=-5pt,
ylabel shift=-5pt
]
\addplot [color=mycolor1, thick, mark size=2pt, mark=square, mark options={solid, mycolor1}]
  table[row sep=crcr]{%
5	1\\
10	1\\
20	0.997\\
30	0.978666666666666\\
50	0.81520000022022\\
};
\addlegendentry{$\text{B=20, P}_{\text{NS}}\text{=0}$}

\addplot [color=mycolor2, thick, dashed, mark size=2pt, mark=o, mark options={solid, mycolor2}]
  table[row sep=crcr]{%
5	1\\
10	0.999\\
20	0.993\\
30	0.970333333333333\\
50	0.7706\\
};
\addlegendentry{$\text{B=20, P}_{\text{NS}}\text{=0.25}$}

\addplot [color=mycolor3, dotted, thick, mark size=2pt, mark=triangle, mark options={solid, mycolor3}]
  table[row sep=crcr]{%
5	0.99\\
10	0.993\\
20	0.9435\\
30	0.881666666666667\\
50	0.6736\\
};
\addlegendentry{$\text{B=20, P}_{\text{NS}}\text{=0.5}$}

\addplot [color=mycolor4, thick]
  table[row sep=crcr]{%
5	1\\
10	1\\
20	1\\
30	1\\
50	0.9984\\
};
\addlegendentry{$\text{B=50, P}_{\text{NS}}\text{=0}$}

\addplot [color=mycolor5, dashed, thick]
  table[row sep=crcr]{%
5	1\\
10	1\\
20	1\\
30	1\\
50	0.9936\\
};
\addlegendentry{$\text{B=50, P}_{\text{NS}}\text{=0.25}$}

\addplot [color=mycolor6, dotted, thick, mark size=2pt]
  table[row sep=crcr]{%
5	0.996\\
10	0.994\\
20	0.994\\
30	0.988666666666667\\
50	0.9684\\
};
\addlegendentry{$\text{B=50, P}_{\text{NS}}\text{=0.5}$}
\end{axis}

\begin{axis}[%
width=0.43\fwidth,
height=0.389\fheight,
at={(0\fwidth,-.05\fheight)},
scale only axis,
xmin=5,
xmax=50,
extra x ticks={50},
extra x tick labels={},
extra x tick style={tick style={draw=none}},
xlabel style={font=\footnotesize\color{white!15!black}},
xlabel={Number of Requests I},
ymin=0,
ymax=1,
ylabel style={font=\footnotesize\color{white!15!black}},
ylabel={Cell site activation ratio},
axis background/.style={fill=white},
axis x line*=bottom,
axis y line*=left,
xmajorgrids,
ymajorgrids,
xlabel shift=-2pt,
ylabel shift=-5pt
]
\addplot [color=mycolor1, thick, mark size=2pt, mark=square, mark options={solid, mycolor1}, forget plot]
  table[row sep=crcr]{%
5	0.2265\\
10	0.393\\
20	0.648\\
30	0.76\\
50	0.7755\\
};
\addplot [color=mycolor2, dashed, thick, mark size=2pt, mark=o, mark options={solid, mycolor2}, forget plot]
  table[row sep=crcr]{%
5	0.2365\\
10	0.4\\
20	0.6535\\
30	0.797\\
50	0.8045\\
};
\addplot [color=mycolor3, dotted, thick, mark size=2pt, mark=triangle, mark options={solid, mycolor3}, forget plot]
  table[row sep=crcr]{%
5	0.235\\
10	0.4155\\
20	0.641\\
30	0.769\\
50	0.8065\\
};
\addplot [color=mycolor4, thick, forget plot]
  table[row sep=crcr]{%
5	0.0972\\
10	0.1876\\
20	0.3284\\
30	0.448\\
50	0.632\\
};
\addplot [color=mycolor5, dashed, thick, forget plot]
  table[row sep=crcr]{%
5	0.0984\\
10	0.1948\\
20	0.3404\\
30	0.4644\\
50	0.6512\\
};
\addplot [color=mycolor6, dotted, thick, forget plot]
  table[row sep=crcr]{%
5	0.0976\\
10	0.1888\\
20	0.3448\\
30	0.4616\\
50	0.6548\\
};
\end{axis}

\begin{axis}[%
width=0.436\fwidth,
height=0.389\fheight,
at={(0.564\fwidth,-.05\fheight)},
scale only axis,
xmin=5,
xmax=50,
extra x ticks={50},
extra x tick labels={},
extra x tick style={tick style={draw=none}},
xlabel style={font=\footnotesize\color{white!15!black}},
xlabel={Number of Requests I},
ymin=0,
ymax=0.3,
ylabel style={font=\footnotesize\color{white!15!black}},
ylabel={Band utilization ratio},
axis background/.style={fill=white},
axis x line*=bottom,
axis y line*=left,
xmajorgrids,
ymajorgrids,
xlabel shift=-2pt,
ylabel shift=-5pt
]
\addplot [color=mycolor1, thick, mark size=2pt, mark=square, mark options={solid, mycolor1}, forget plot]
  table[row sep=crcr]{%
5	0.03795\\
10	0.0743\\
20	0.14905\\
30	0.22065\\
50	0.2829\\
};
\addplot [color=mycolor2, dashed, thick, mark size=2pt, mark=o, mark options={solid, mycolor2}, forget plot]
  table[row sep=crcr]{%
5	0.0372\\
10	0.0745\\
20	0.14825\\
30	0.21795\\
50	0.2625\\
};
\addplot [color=mycolor3, dotted, thick, mark size=2pt, mark=triangle, mark options={solid, mycolor3}, forget plot]
  table[row sep=crcr]{%
5	0.0378\\
10	0.07465\\
20	0.1402\\
30	0.19515\\
50	0.2142\\
};
\addplot [color=mycolor4, thick, forget plot]
  table[row sep=crcr]{%
5	0.01496\\
10	0.03008\\
20	0.05924\\
30	0.09048\\
50	0.15104\\
};
\addplot [color=mycolor5, dashed, thick, forget plot]
  table[row sep=crcr]{%
5	0.01516\\
10	0.03052\\
20	0.0594\\
30	0.089\\
50	0.15024\\
};
\addplot [color=mycolor6, dotted, thick, forget plot]
  table[row sep=crcr]{%
5	0.0146\\
10	0.03\\
20	0.06048\\
30	0.08848\\
50	0.14512\\
};
\end{axis}
\end{tikzpicture}%
        \setlength\abovecaptionskip{0cm}
    \setlength\belowcaptionskip{-.45cm}
    \caption{Acceptance, cell site activation and band utilization ratios as a function of the number of requests ($I$) for different values of $P_{\mr{NS}}$ and number of cell sites ($B$). The figures share the same legend. $W=5$.}
    \label{fig:pns}
\end{figure}

Figure~\ref{fig:pns} depicts the acceptance, cell site activation, and band utilization ratios as a function of the number of requests for different values of $P_{\mr{NS}}$ and number $B$ of cell sites controlled by \neutran. 
The figure on the top shows that the acceptance ratio decreases when a larger number of requests is submitted to \neutran. Intuitively, the larger the number of requests, the higher the probability that not all requests can be accommodated due to the limited number of \prbs and spectrum bands. 
Moreover, when $B=20$, the acceptance ratio decreases from 82\% for $P_{\mr{NS}}=0$ (i.e., all cell sites support all $W=5$ bands) to 67\% for $P_{\mr{NS}}=0.5$ (i.e., the probability that a cell site does not support a band is 50\%).
%
The same applies when $B=50$, with the acceptance ratio dropping from 99.8\% for $P_{\mr{NS}}=0$ to 96\% for $P_{\mr{NS}}=0.5$. As expected, more cell sites also imply a higher acceptance rate as more resources are available to \neutran.

Figure~\ref{fig:pns} (bottom-left) shows that the cell site activation ratio always increases with the number of requests submitted to \neutran. 
The activation ratio is higher for $B=20$ than $B=50$, as \neutran's optimal policy requires the allocation of as many cell sites as possible in the available pool.
%
Instead, Figure~\ref{fig:pns} (bottom-right) shows how different configurations affect spectrum utilization. Intuitively, the higher the number of submitted requests, the higher the number of spectrum bands to be allocated to support such requests and avoid interference among neighboring cell sites. Indeed, as the cell site activation ratio increases, more bands must be activated to eliminate interference while allocating the necessary \prbs. 
This happens in particular for the resource constrained case of $B=20$.


\ifexttikz
    \tikzsetnextfilename{psb}
\fi
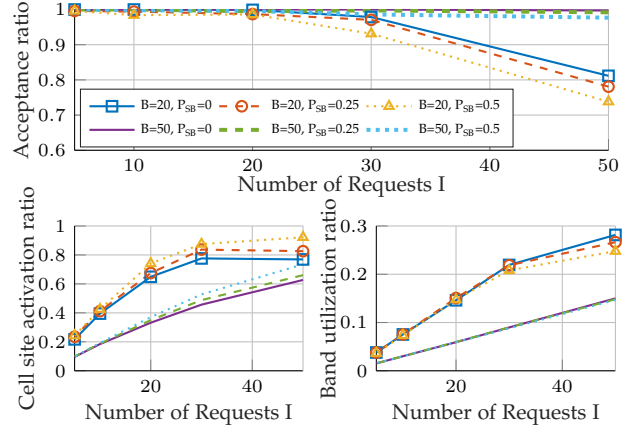
\begin{figure}[t!]
    \centering
    \setlength\fwidth{0.8\columnwidth}
    \setlength\fheight{.55\columnwidth}
%
%
\definecolor{mycolor1}{rgb}{0.00000,0.44700,0.74100}%
\definecolor{mycolor2}{rgb}{0.85000,0.32500,0.09800}%
\definecolor{mycolor3}{rgb}{0.92900,0.69400,0.12500}%
\definecolor{mycolor4}{rgb}{0.49400,0.18400,0.55600}%
\definecolor{mycolor5}{rgb}{0.46600,0.67400,0.18800}%
\definecolor{mycolor6}{rgb}{0.30100,0.74500,0.93300}%
\begin{tikzpicture}
\pgfplotsset{every tick label/.append style={font=\scriptsize}}

\begin{axis}[%
width=0.998\fwidth,
height=0.383\fheight,
at={(0\fwidth,0.55\fheight)},
scale only axis,
xmin=5,
xmax=50,
xlabel style={font=\footnotesize\color{white!15!black}},
xlabel={Number of Requests I},
ymin=0.6,
ymax=1,
ylabel style={font=\footnotesize\color{white!15!black}},
ylabel={Acceptance ratio},
axis background/.style={fill=white},
axis x line*=bottom,
axis y line*=left,
xmajorgrids,
ymajorgrids,
legend style={at={(0.01, 0.02)}, anchor=south west, legend cell align=left, align=left, draw=white!15!black, font=\tiny},
legend columns=3,
xlabel shift=-5pt,
ylabel shift=-5pt
]
\addplot [color=mycolor1, thick, mark size=2pt, mark=square, mark options={solid, mycolor1}]
  table[row sep=crcr]{%
50	0.811600000066667\\
30	0.978666666666667\\
20	0.999\\
10	1\\
5	1\\
};
\addlegendentry{$\text{B=20, P}_{\text{SB}}\text{=0}$}

\addplot [color=mycolor2, dashed, thick, mark size=2pt, mark=o, mark options={solid, mycolor2}]
  table[row sep=crcr]{%
50	0.7808\\
30	0.970666666666667\\
20	0.987\\
10	0.994\\
5	0.996\\
};
\addlegendentry{$\text{B=20, P}_{\text{SB}}\text{=0.25}$}

\addplot [color=mycolor3, dotted, thick, mark size=2pt, mark=triangle, mark options={solid, mycolor3}]
  table[row sep=crcr]{%
50	0.738\\
30	0.931333333333333\\
20	0.986\\
10	0.984\\
5	0.996\\
};
\addlegendentry{$\text{B=20, P}_{\text{SB}}\text{=0.5}$}

\addplot [color=mycolor4, thick]
  table[row sep=crcr]{%
50	0.9972\\
30	1\\
20	1\\
10	1\\
5	1\\
};
\addlegendentry{$\text{B=50, P}_{\text{SB}}\text{=0}$}

\addplot [color=mycolor5, dashed, thick, line width=1.5pt]
  table[row sep=crcr]{%
50	0.992\\
30	0.997333333333333\\
20	0.995\\
10	1\\
5	1\\
};
\addlegendentry{$\text{B=50, P}_{\text{SB}}\text{=0.25}$}

\addplot [color=mycolor6, dotted, thick, line width=1.5pt]
  table[row sep=crcr]{%
50	0.9768\\
30	0.986\\
20	0.996\\
10	0.998\\
5	1\\
};
\addlegendentry{$\text{B=50, P}_{\text{SB}}\text{=0.5}$}

\end{axis}

\begin{axis}[%
width=0.427\fwidth,
height=0.394\fheight,
at={(0\fwidth,-0.05\fheight)},
scale only axis,
xmin=5,
xmax=50,
extra x ticks={50},
extra x tick labels={},
extra x tick style={tick style={draw=none}},
xlabel style={font=\footnotesize\color{white!15!black}},
xlabel={Number of Requests I},
ymin=0,
ymax=1,
ylabel style={font=\footnotesize\color{white!15!black}},
ylabel={Cell site activation ratio},
axis background/.style={fill=white},
axis x line*=bottom,
axis y line*=left,
xmajorgrids,
ymajorgrids,
xlabel shift=-2pt,
ylabel shift=-5pt
]
\addplot [color=mycolor1, thick, mark size=2pt, mark=square, mark options={solid, mycolor1}, forget plot]
  table[row sep=crcr]{%
50	0.769\\
30	0.776\\
20	0.649\\
10	0.395\\
5	0.217\\
};
\addplot [color=mycolor2, dashed, thick, mark size=2pt, mark=o, mark options={solid, mycolor2}, forget plot]
  table[row sep=crcr]{%
50	0.826\\
30	0.837\\
20	0.675\\
10	0.412\\
5	0.235\\
};
\addplot [color=mycolor3, dotted, thick, mark size=2pt, mark=triangle, mark options={solid, mycolor3}, forget plot]
  table[row sep=crcr]{%
50	0.922\\
30	0.875\\
20	0.741\\
10	0.425\\
5	0.243\\
};
\addplot [color=mycolor4, thick, forget plot]
  table[row sep=crcr]{%
50	0.6272\\
30	0.4568\\
20	0.332\\
10	0.1844\\
5	0.096\\
};
\addplot [color=mycolor5, dashed, thick, forget plot]
  table[row sep=crcr]{%
50	0.6604\\
30	0.488\\
20	0.3496\\
10	0.1876\\
5	0.0988\\
};
\addplot [color=mycolor6, dotted, thick, forget plot]
  table[row sep=crcr]{%
50	0.7352\\
30	0.5276\\
20	0.3704\\
10	0.1924\\
5	0.0984\\
};
\end{axis}

\begin{axis}[%
width=0.447\fwidth,
height=0.394\fheight,
at={(0.564\fwidth,-0.05\fheight)},
scale only axis,
xmin=5,
xmax=50,
extra x ticks={50},
extra x tick labels={},
extra x tick style={tick style={draw=none}},
xlabel style={font=\footnotesize\color{white!15!black}},
xlabel={Number of Requests I},
ymin=0,
ymax=0.3,
ylabel style={font=\footnotesize\color{white!15!black}},
ylabel={Band utilization ratio},
axis background/.style={fill=white},
axis x line*=bottom,
axis y line*=left,
xmajorgrids,
ymajorgrids,
xlabel shift=-2pt,
ylabel shift=-5pt
]
\addplot [color=mycolor1, thick, mark size=2pt, mark=square, mark options={solid, mycolor1}, forget plot]
  table[row sep=crcr]{%
50	0.2815\\
30	0.2193\\
20	0.1461\\
10	0.0753\\
5	0.0379\\
};
\addplot [color=mycolor2, dashed, thick, mark size=2pt, mark=o, mark options={solid, mycolor2}, forget plot]
  table[row sep=crcr]{%
50	0.2664\\
30	0.2186\\
20	0.1507\\
10	0.0753\\
5	0.0378\\
};
\addplot [color=mycolor3, dotted, thick, mark size=2pt, mark=triangle, mark options={solid, mycolor3}, forget plot]
  table[row sep=crcr]{%
50	0.2479\\
30	0.208\\
20	0.1475\\
10	0.0735\\
5	0.0364\\
};
\addplot [color=mycolor4, thick, forget plot]
  table[row sep=crcr]{%
50	0.15\\
30	0.0898\\
20	0.0592\\
10	0.03016\\
5	0.01524\\
};
\addplot [color=mycolor5, dashed, thick, forget plot]
  table[row sep=crcr]{%
50	0.14824\\
30	0.08988\\
20	0.0598\\
10	0.02984\\
5	0.01552\\
};
\addplot [color=mycolor6, dotted, thick, forget plot]
  table[row sep=crcr]{%
50	0.14684\\
30	0.0878\\
20	0.0592\\
10	0.02988\\
5	0.01568\\
};
\end{axis}
\end{tikzpicture}%
    \setlength\belowcaptionskip{-.3cm}
    \setlength\abovecaptionskip{0cm}
    \caption{Acceptance, cell site activation and band utilization ratios as a function of the number of requests ($I$) for different values of $P_{\mr{SB}}$ and number of cell sites ($B$). The figures share the same legend. $W=5$.}
    \label{fig:psb}
\end{figure}

Figure~\ref{fig:psb} concerns the same metrics 
for varying values of the probability $P_{\mr{SB}}$. Trends for acceptance and band utilization rates are similar to those in Figure~\ref{fig:pns}. However, we notice that varying $P_{\mr{SB}}$ affects the activation of cell sites to a higher extent (Figure~\ref{fig:psb}, bottom-left). Specifically, we see that when half the cell sites can operate on a single spectrum band at any given time ($P_{\mr{SB}}\!=\!0.5$), the activation ratio increases by approximately 20\% compared to the case when all cell sites support multiple spectrum bands ($P_{\mr{SB}}\!=\!0$). This illustrates the importance of deploying cell sites with \rus that can transmit simultaneously on multiple bands and that can accommodate more requests with a lower number of active cell sites (lowering operational and energy costs).

\begin{figure}[b]
  \setlength\abovecaptionskip{2pt}
  \centering
  \includegraphics[width=.95\columnwidth]{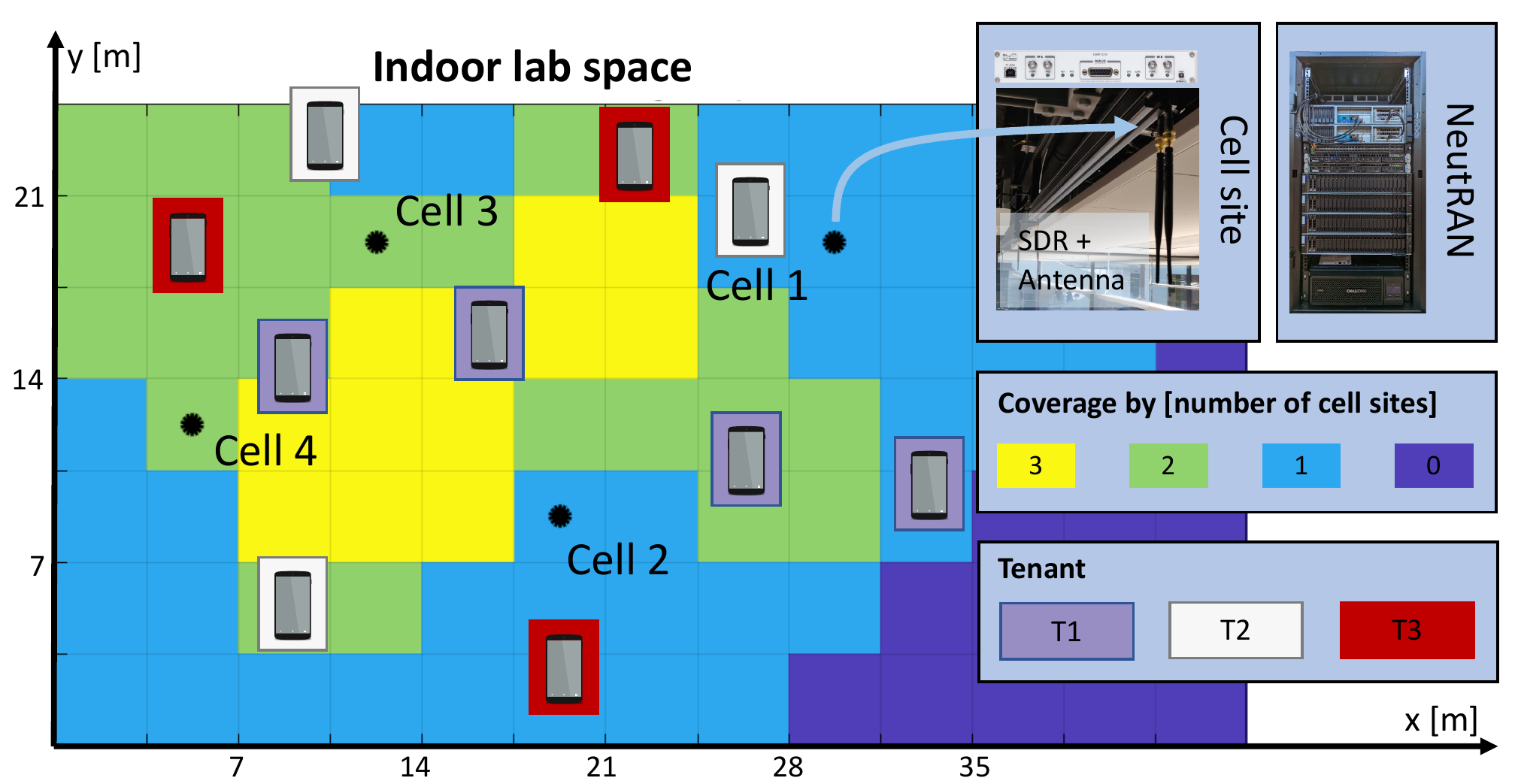}
  \caption{Experimental setup: \neutran on an over-the-air \gls{sdr} testbed.}
  \label{fig:exp-setup}
\end{figure}

\setcounter{figure}{8}
\begin{figure*}[b]
  \centering
  \vspace{-.3cm}
  \ifexttikz
    \tikzsetnextfilename{throughput-cdf}
  \fi
  \begin{subfigure}[t]{0.32\textwidth}
    \setlength\fwidth{0.85\columnwidth}
    \setlength\fheight{0.5\columnwidth}
        \includegraphics[width=\columnwidth]{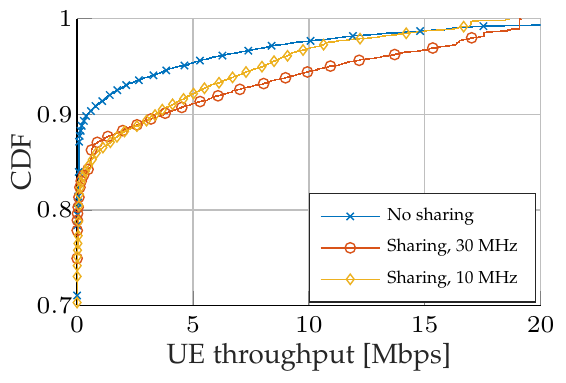}
    \setlength\abovecaptionskip{-.4cm}
    \caption{CDF of the UE throughput}
    \label{fig:throughput-cdf}
  \end{subfigure}
  \ifexttikz
      \tikzsetnextfilename{dl-mcs-cdf}
  \fi
  \begin{subfigure}[t]{0.32\textwidth}
    \setlength\fwidth{0.85\columnwidth}
    \setlength\fheight{0.5\columnwidth}
        \includegraphics[width=\columnwidth]{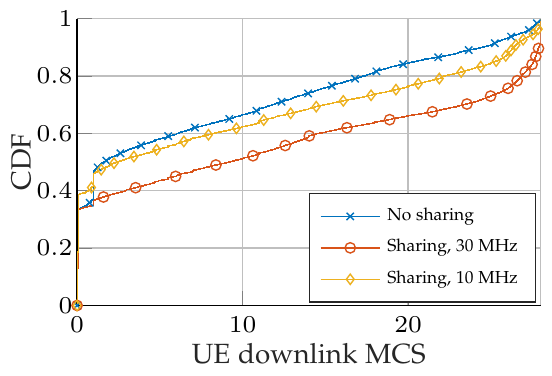} 
    \setlength\abovecaptionskip{-0.4cm}
    \caption{CDF of the UE downlink MCS}
    \label{fig:mcs-cdf}
  \end{subfigure}
  \ifexttikz
      \tikzsetnextfilename{ul-sinr-cdf}
  \fi
  \begin{subfigure}[t]{0.32\textwidth}
    \setlength\fwidth{0.85\columnwidth}
    \setlength\fheight{0.5\columnwidth}
        \includegraphics[width=\columnwidth]{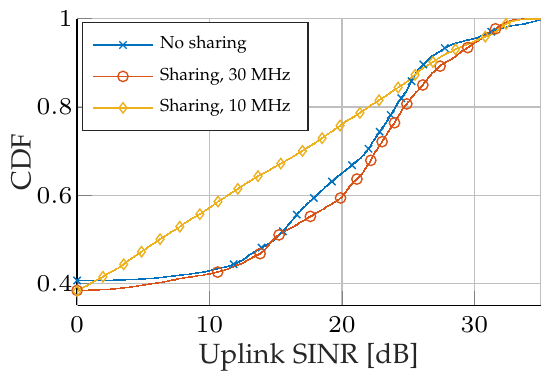}
    \setlength\abovecaptionskip{-0.4cm}
    \caption{CDF of the uplink SINR}
    \label{fig:sinr-cdf}
  \end{subfigure}
  \setlength\belowcaptionskip{-.3cm}
  \caption{CDF of UE throughput, downlink MCS, and uplink SINR. With \emph{No sharing}, three tenants deploy their RAN on independent spectrum ($10$\:MHz each, for a total of $30$\:MHz) and cell sites and equipment. With \emph{Sharing, $30$\:MHz}, the tenants share the RAN infrastructure, cell sites, and $30$\:MHz of spectrum with the \neutran architecture. Finally, \emph{Sharing, $10$\:MHz} features the same configuration, but only uses $10$\:MHz of spectrum.}
  \label{fig:cdf}
\end{figure*}





\subsection{Experimental Evaluation of \ran and Spectrum Sharing}
\label{sec:exp-results}

To evaluate the performance of \neutran in a real-world scenario, we deployed the prototype in an indoor space with multiple obstacles, heterogeneous equipment and moving humans, creating a wireless environment rich with scattering.
The indoor area (Figure~\ref{fig:exp-setup}) covers more than $1100$\:m$^2$. It has been logically divided into $A=91$ tiles of size $3.5 \times 3.5$\:m$^2$.
We consider a configuration with $B=4$ cell sites, each with an USRP X310 frontend, whose antennas are mounted on the ceiling of the testbed.
We deployed 10 commercial smartphones as shown in Figure~\ref{fig:exp-setup}.
Smartphones represent end users for a set $\mathcal{T}$ with $T=3$ different tenants (e.g., different mobile network or private network operators). Specifically, 4 users are served by tenant 1, and 3 each by tenants 2 and 3. Connectivity is provided over LTE Band 7 under an experimental license, and $W\!=\!3$ blocks with $10$\:MHz of spectrum  (50~\glspl{prb}) each. Blocks are identified by a pair $(f_d,f_u)$ of downlink and uplink carrier frequencies, with $f_d \in \{2.625, 2.645, 2.685\}$\:GHz and $ f_u \in \{2.505, 2.525, 2.565\}$\:GHz.
We conservatively define the coverage area of a cell site as the set of tiles where users experience an average throughput higher than 50\% of the throughput that they would experience in the tile with the cell site antenna. The result is the coverage map in Figure~\ref{fig:exp-setup}.

To exhaustively evaluate the performance of \neutran, we consider the three following scenarios, each involving four independent experiments lasting more than $300$\:s each.

\noindent$\bullet$~\emph{No sharing:} this license-based scenario corresponds to a traditional cellular network deployment where each tenant (i)~uses different cell sites without infrastructure sharing, and~(ii) owns a dedicated, licensed portion of the spectrum. Each tenant operates independently on one of the three $10$\:MHz blocks. Since the \ran is not shared, tenant~1 serves users from cell sites~1 and~3, tenant~2 from cell site~2, and tenant~3 from cell site~4.

\noindent$\bullet$~\emph{Sharing, $30$\:MHz:} 
the \neutran optimization engine and automation pipelines are used to allocate tenant requests over different cell sites and spectrum. \neutran controls a total of $30$\:MHz, i.e., the same amount of spectrum available to the \emph{No sharing} configuration. Each cell site, however, only provides service through a single pair of downlink/uplink carrier frequencies and a $10$\:MHz spectrum block.

\noindent$\bullet$~\emph{Sharing, $10$\:MHz:} 
this configuration is the same as \emph{Sharing, $30$\:MHz}, with the difference that the system is deployed on an overall spectrum of $10$\:MHz, shared among tenants.
Since only one band is available in this scenario ($W=1$), \neutran optimization engine does not enforce the interference constraint~\eqref{eq:con:zero_intf}.

The \glspl{kpm} are collected at the \gls{ran} side every $250$\:ms for each \gls{ue}. Here, we consider downlink throughput, downlink \gls{mcs}, and uplink \gls{sinr}. We also aggregate the \gls{ue} downlink throughput over time to compute the average and median \gls{ue} throughput and the total \gls{ran} throughput for each experiment. During the experiments, the smartphones stream videos through the YouTube application. 

\setcounter{figure}{7}
\ifexttikz
    \tikzsetnextfilename{throughput-bar}
\fi
\begin{figure}[t]
  \centering
  \setlength\fwidth{0.9\columnwidth}
  \setlength\fheight{0.5\columnwidth}
%
%
\definecolor{mycolor1}{rgb}{0.00000,0.44700,0.74100}%
\definecolor{mycolor2}{rgb}{0.85000,0.32500,0.09800}%
\definecolor{mycolor3}{rgb}{0.92900,0.69400,0.12500}%
\begin{tikzpicture}
\pgfplotsset{every tick label/.append style={font=\footnotesize}}

\begin{axis}[%
width=0.951\fwidth,
height=\fheight,
at={(0\fwidth,0\fheight)},
scale only axis,
bar shift auto,
xmin=0.511111111111111,
xmax=2.48888888888889,
extra x ticks={2.48888888888889},
extra x tick labels={},
extra x tick style={tick style={draw=none}},
xtick={1,2},
xticklabels={{Average UE throughput},{Total RAN throughput}},
ymin=0,
ymax=18,
extra y ticks={18},
extra y tick labels={},
extra y tick style={tick style={draw=none}},
ylabel style={font=\footnotesize\color{white!15!black}},
xlabel style={font=\footnotesize\color{white!15!black}},
ylabel={Throughput [Mbps]},
axis background/.style={fill=white},
axis x line*=bottom,
axis y line*=left,
xmajorgrids,
ymajorgrids,
legend style={legend cell align=left, align=left, draw=white!15!black, font=\tiny, at={(1, 0)}, anchor=south east}
]
\addplot[ybar, bar width=0.178, fill=mycolor1, draw=black, area legend] table[row sep=crcr] {%
1	0.695240953487332\\
2	5.90954810464232\\
};
\addplot[forget plot, color=white!15!black] table[row sep=crcr] {%
0.511111111111111	0\\
2.48888888888889	0\\
};
\addlegendentry{No sharing}

\addplot[ybar, bar width=0.178, fill=mycolor2, draw=black, area legend] table[row sep=crcr] {%
1	1.20254146557729\\
2	12.9273207549559\\
};
\addplot[forget plot, color=white!15!black] table[row sep=crcr] {%
0.511111111111111	0\\
2.48888888888889	0\\
};
\addlegendentry{Sharing, 30 MHz}

\addplot[ybar, bar width=0.178, fill=mycolor3, draw=black, area legend] table[row sep=crcr] {%
1	0.873288909232071\\
2	10.4794669107848\\
};
\addplot[forget plot, color=white!15!black] table[row sep=crcr] {%
0.511111111111111	0\\
2.48888888888889	0\\
};
\addlegendentry{Sharing, 10 MHz}

\addplot [color=black, only marks, mark size=0pt, forget plot]
 plot [error bars/.cd, y dir=both, y explicit, error bar style={line width=0.5pt}, error mark options={line width=0.5pt, mark size=3.0pt, rotate=90}]
 table[row sep=crcr, y error plus index=2, y error minus index=3]{%
0.805	0.695240953487332	0.248984638298629	0.248984638298629\\
1.805	5.90954810464232	3.16412529374191	3.16412529374191\\
};
\addplot [color=black, only marks,  mark size=0pt, forget plot]
 plot [error bars/.cd, y dir=both, y explicit, error bar style={line width=0.5pt}, error mark options={line width=0.5pt, mark size=3.0pt, rotate=90}]
 table[row sep=crcr, y error plus index=2, y error minus index=3]{%
1	1.20254146557729	0.345454722084482	0.345454722084482\\
2	12.9273207549559	4.94506764386359	4.94506764386359\\
};
\addplot [color=black, only marks,  mark size=0pt, forget plot]
 plot [error bars/.cd, y dir=both, y explicit, error bar style={line width=0.5pt}, error mark options={line width=0.5pt, mark size=3.0pt, rotate=90}]
 table[row sep=crcr, y error plus index=2, y error minus index=3]{%
1.195	0.873288909232071	0.300349558042565	0.300349558042565\\
2.195	10.4794669107848	2.68396345932821	2.68396345932821\\
};
\end{axis}

\node [above left,
       fill=white,
       align=center,
       anchor=north,
       font=\tiny] at (2.7,4.25) {%
    \tiny
    \setlength{\tabcolsep}{3pt}
    \begin{tabularx}{0.56\columnwidth}{
        >{\hsize=1.3\hsize}X
        >{\hsize=0.9\hsize}X
        >{\hsize=0.9\hsize}X
        >{\hsize=0.9\hsize}X}
    \midrule 
    & Average~UE & Median~UE & RAN \\
    & \multicolumn{3}{c}{throughput [Mbps]}\\\midrule 
    No sharing & 0.6952  & 0.0827 &  5.9095  \\
    Sharing,\,30\:MHz & 1.2025  & 0.6065 & 12.9273  \\
    Sharing,\,10\:MHz & 0.8733  & 0.0820 & 10.4795  \\
    \midrule 
    \end{tabularx}

    };

\end{tikzpicture}%
  \setlength\belowcaptionskip{-.3cm}
  \caption{Average, sum, and median throughput metrics. With \emph{No sharing}, three tenants deploy their RAN on independent spectrum ($10$\:MHz each, for a total of $30$\:MHz) and cell sites and equipment. With \emph{Sharing, $30$\:MHz}, the tenants share the RAN infrastructure, cell sites, and $30$\:MHz of spectrum with the \neutran architecture. Finally, \emph{Sharing, $10$\:MHz} features the same sharing configuration, but only uses $10$\:MHz of spectrum. We show the 95\% confidence intervals.}
  \label{fig:throughput}
\end{figure}

\subsubsection{Experimental Results}
\label{sec:results}

We investigate whether using the \neutran-enabled \gls{ran} and spectrum sharing improves network performance. 
Figure~\ref{fig:throughput} concerns the average throughput. 
%

%
Both \emph{Sharing} configurations outperform the first one in the total \gls{ran} throughput, showing that a network with an optimized approach to resource management and sharing outperforms orthogonal license-based schemes.
\emph{Sharing, $30$\:MHz} achieves a 2.18$\times$ gain over \emph{No Sharing}, and the setup with $10$\:MHz spectrum has a 1.77$\times$ gain, as also shown in the table part of Figure~\ref{fig:throughput}. In addition, the \emph{Sharing} approach with $30$\:MHz also outperforms the \emph{No Sharing} scheme in terms of average and median throughput, suggesting that the improvement does not apply only to aggregated performance, but also, on average, to the single \glspl{ue}.
%
A similar behavior can be observed in Figures~\ref{fig:throughput-cdf}-\ref{fig:mcs-cdf},
which show the \gls{cdf} of the \gls{ue} throughput and downlink \gls{mcs}, respectively. Both \textit{Sharing} scenarios outperform the \textit{No Sharing} one.

Considering the \textit{Sharing} setup with $10$\:MHz, we notice that sharing the available spectrum improves the total throughput when compared to the case \textit{No sharing},
with comparable average and median \gls{ue} throughput.
%
%
We would like to point out that, thanks to the capabilities enabled by \neutran, such comparable behavior is achieved by using $1/3$ of the spectrum used in the \textit{No sharing} setup. This is because \gls{ran} sharing allows tenants to optimally deploy their resources, e.g., by maximizing the coverage of their users and thus improving the downlink throughput and \gls{mcs} (Figures~\ref{fig:throughput-cdf}-\ref{fig:mcs-cdf}), especially for users with the best channel conditions.
%
%
%
However, when using a single 10 MHz block, interference is significant for cell edge users, as shown in Figure~\ref{fig:sinr-cdf}, which reports the uplink \gls{sinr} experienced by each \gls{ue} and shows an average 5~dB gap for 80\% of the \glspl{ue} between the $10$\:MHz \textit{Sharing} setup and the others.
The linear behavior observed in this scenario is due to the higher spectrum utilization of the three tenants sharing a narrower portion of the spectrum. This affects the \gls{pdf} of \gls{mcs} and \gls{sinr}, reported in Figure~\ref{fig:ul_mcs_sinr}, with values close to those of the uniform distribution in the $10$\:MHz \textit{Sharing} setup.
%
%
%
This causes less fairness among the \glspl{ue} in uplink.

\setcounter{figure}{9}
\ifexttikz
    \tikzsetnextfilename{ul_mcs_sinr}
\fi
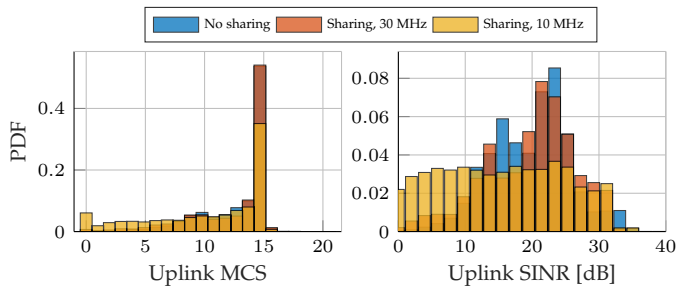
\begin{figure}[t!]
    \centering
    \setlength\fwidth{\columnwidth}
    \setlength\fheight{.55\columnwidth}
%
%
\definecolor{mycolor1}{rgb}{0.00000,0.44700,0.74100}%
\definecolor{mycolor2}{rgb}{0.85000,0.32500,0.09800}%
\definecolor{mycolor3}{rgb}{0.92900,0.69400,0.12500}%
\begin{tikzpicture}
\pgfplotsset{every tick label/.append style={font=\scriptsize}}

\begin{axis}[%
width=0.4\fwidth,
height=0.49\fheight,
at={(0.485\fwidth,0\fheight)},
scale only axis,
xmin=0,
xmax=40,
extra x ticks={40},
extra x tick labels={},
extra x tick style={tick style={draw=none}},
ylabel style={font=\footnotesize\color{white!15!black}},
xlabel style={font=\footnotesize\color{white!15!black}},
xlabel={Uplink SINR [dB]},
axis background/.style={fill=white},
axis x line*=bottom,
axis y line*=left,
xmajorgrids,
ymajorgrids,
ymin=0,
ymax=0.094,
extra y ticks={0.094},
extra y tick labels={},
extra y tick style={tick style={draw=none}},
ylabel shift=-2pt,
xlabel shift=-2pt,
yticklabel style={/pgf/number format/fixed, /pgf/number format/precision=5},
scaled y ticks=false,
legend style={legend cell align=left, align=left, draw=white!15!black, font=\tiny, at={(1, 0)}, anchor=south east}
]
\addplot[ybar, bar width=1.75, fill=mycolor1, opacity=0.75, draw=black, area legend] table[row sep=crcr] {%
x	y\\
0	0.000553116128671961\\
1.947685	0.000967953225175932\\
3.89537	0.00228160403077184\\
5.843055	0.00414837096503971\\
7.79074	0.00691395160839951\\
9.738425	0.014726716925891\\
11.68611	0.0335326653007376\\
13.633795	0.0407231749734731\\
15.58148	0.0588377281874798\\
17.529165	0.0463234757762767\\
19.47685	0.0408614540056411\\
21.424535	0.0729421894686148\\
23.37222	0.085387302363734\\
25.319905	0.0508866838378204\\
27.26759	0.0206727153091145\\
29.215275	0.0102326483804313\\
31.16296	0.0107166249930192\\
33.110645	0.0109931830573552\\
35.05833	0.00165934838601588\\
37.006015	0\\
38.9537	6.91395160839951e-05\\
40.901385	6.91395160839951e-05\\
};

\addplot[ybar, bar width=1.75, fill=mycolor2, opacity=0.75, draw=black, area legend] table[row sep=crcr] {%
x	y\\
0	0.00205137575632319\\
1.947685	0.00549475649015141\\
3.89537	0.00835202986503014\\
5.843055	0.00901140064384831\\
7.79074	0.00893813722397963\\
9.738425	0.018169328127434\\
11.68611	0.0277668361302318\\
13.633795	0.045643110578191\\
15.58148	0.0279133629699692\\
17.529165	0.0306973729249792\\
19.47685	0.052163554946504\\
21.424535	0.0783185958396248\\
23.37222	0.070332883073938\\
25.319905	0.0508448133888677\\
27.26759	0.0292321045276055\\
29.215275	0.0255689335341712\\
31.16296	0.0214661820215248\\
33.110645	0.00139200497750502\\
35.05833	7.32634198686855e-05\\
37.006015	7.32634198686855e-05\\
};

\addplot[ybar, bar width=1.75, fill=mycolor3, opacity=0.75, draw=black, area legend] table[row sep=crcr] {%
x	y\\
0	0.0219976149431675\\
1.947685	0.0287397410815928\\
3.89537	0.0308537975826245\\
5.843055	0.0329678540836562\\
7.79074	0.0321108041508055\\
9.738425	0.0335963573677467\\
11.68611	0.0320536674886155\\
13.633795	0.0295396543522535\\
15.58148	0.0309680709070046\\
17.529165	0.034053450665267\\
19.47685	0.0322822141373756\\
21.424535	0.0324536241239458\\
23.37222	0.0366817371260091\\
25.319905	0.0336534940299367\\
27.26759	0.0232546215113485\\
29.215275	0.0212548383346969\\
31.16296	0.0250258580392399\\
33.110645	0.00194264651446154\\
35.05833	0.00194264651446154\\
};

\end{axis}

\begin{axis}[%
width=0.4\fwidth,
height=0.49\fheight,
at={(0\fwidth,0\fheight)},
scale only axis,
xmin=-1.02725175,
xmax=21.57228675,
extra x ticks={21.57228675},
extra x tick labels={},
extra x tick style={tick style={draw=none}},
xtick={0,5,10,15,20},
ylabel style={font=\footnotesize\color{white!15!black}},
ylabel={PDF},
xlabel style={font=\footnotesize\color{white!15!black}},
xlabel={Uplink MCS},
axis background/.style={fill=white},
ymin=0,
ymax=0.585,
extra y ticks={0.585},
extra y tick labels={},
extra y tick style={tick style={draw=none}},
axis x line*=bottom,
axis y line*=left,
xmajorgrids,
ymajorgrids,
ylabel shift=-2pt,
xlabel shift=-2pt,
legend columns=3,
legend style={legend cell align=left, align=left, draw=white!15!black, font=\tiny, at={(1.11, 1.05)}, anchor=south}
]
\addplot[ybar, bar width=1, fill=mycolor1, opacity=0.75, draw=black, area legend] table[row sep=crcr] {%
x	y\\
0	0.00137644056876281\\
0.978335	0.000688220284381403\\
1.95667	0.00151408462563909\\
2.935005	0.00178937273939165\\
3.91334	0.00633162661630891\\
4.891675	0.00660691473006147\\
5.87001	0.0134891175738755\\
6.848345	0.0181690155076691\\
7.82668	0.0341357261053176\\
8.805015	0.0516165213286053\\
9.78335	0.0622151137080789\\
10.761685	0.0462484031104303\\
11.74002	0.0551952668073886\\
12.718355	0.0776312480782222\\
13.69669	0.0991037209509221\\
14.675025	0.536674177760618\\
15.65336	0.00908450775383453\\
16.631695	0.000275288113752561\\
17.61003	0.000275288113752561\\
};
\addlegendentry{No sharing}

\addplot[ybar, bar width=1, fill=mycolor2, opacity=0.75, draw=black, area legend] table[row sep=crcr] {%
x	y\\
0	0.00685513755979926\\
0.978335	0.00350049577521664\\
1.95667	0.00495903568155691\\
2.935005	0.00889709342867564\\
3.91334	0.00831367746613953\\
4.891675	0.0128351511757944\\
5.87001	0.0211488286419339\\
6.848345	0.0245034704265165\\
7.82668	0.032671293902022\\
8.805015	0.0538201225439559\\
9.78335	0.0545493924971261\\
10.761685	0.0398181394430893\\
11.74002	0.043172781227672\\
12.718355	0.0503196267687392\\
13.69669	0.102535355415721\\
14.675025	0.540826597270972\\
15.65336	0.0129810051664284\\
16.631695	0.00043756197190208\\
17.61003	0.00043756197190208\\
};
\addlegendentry{Sharing, 30 MHz}

\addplot[ybar, bar width=1, fill=mycolor3, opacity=0.75, draw=black, area legend] table[row sep=crcr] {%
x	y\\
0	0.0606279947108411\\
0.978335	0.0189960133521772\\
1.95667	0.0290058886515281\\
2.935005	0.0328733404717318\\
3.91334	0.0333283348035205\\
4.891675	0.0312808603104715\\
5.87001	0.035717055045411\\
6.848345	0.0377645295384601\\
7.82668	0.0367407922919356\\
8.805015	0.0456131817618148\\
9.78335	0.0507318679944374\\
10.761685	0.0483431477525468\\
11.74002	0.0536893311510638\\
12.718355	0.0685903955171428\\
13.69669	0.0798515052289126\\
14.675025	0.350914378392016\\
15.65336	0.00727990930861882\\
16.631695	0.000454994331788675\\
17.61003	0.000113748582947169\\
18.588365	0.000113748582947169\\
19.5667	0.000113748582947169\\
20.545035	0.000113748582947169\\
};
\addlegendentry{Sharing, 10 MHz}

\end{axis}

\end{tikzpicture}%
    \caption{PDF of UE uplink MCS and SINR in the \emph{No sharing}, \emph{Sharing, $30$\:MHz}, and \emph{Sharing, $10$\:MHz} scenarios. The MCS saturates to~15 for values of SINR $\ge 22.7$.}
    \label{fig:ul_mcs_sinr}
\end{figure}

%
%

Another aspect worth considering is related to how fast \neutran can deploy a fully working cellular network. Table~\ref{tab:inst-time} shows the average instantiation time for the pods of specific micro-services (i.e., base station, core network, near-RT \gls{ric}, and xApp pods), as well as the number of pods required to run them.
\begin{table}[t]
\setlength\belowcaptionskip{5pt}
    \centering
    \footnotesize
    \setlength{\tabcolsep}{2pt}
    \caption{Average deployment and instantiation time of different applications on the OpenShift-managed \neutran infrastructure.}
    \label{tab:inst-time}
    \begin{tabularx}{0.95\columnwidth}{
        >{\raggedright\arraybackslash\hsize=\hsize}X
        >{\raggedleft\arraybackslash\hsize=\hsize}X
        >{\raggedleft\arraybackslash\hsize=\hsize}X }
        \toprule
        Application & Required Pods & Instantiation Time [s] \\
        \midrule
        Base station    & 1         & 9.55 \\
        Core network    & 17        & 5.90 \\
        Near-RT RIC     & 13        & 2.93 \\
        xApp            & 1         & 2.35 \\
        \bottomrule
    \end{tabularx}
\end{table}

\ifexttikz
    \tikzsetnextfilename{pod-times}
\fi
\begin{figure}[ht]
    \setlength\belowcaptionskip{0pt}
    \centering
    \setlength\fwidth{0.9\columnwidth}
    \setlength\fheight{.32\columnwidth}
    \input{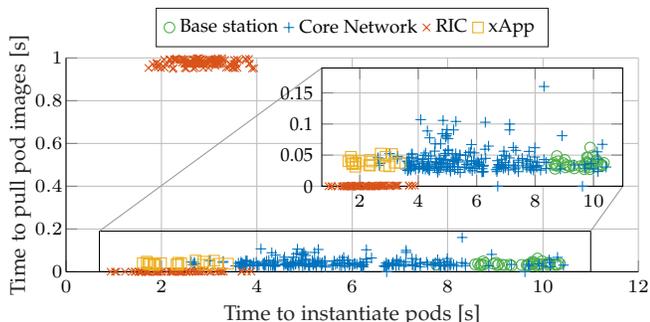}
    \setlength\abovecaptionskip{0cm}
    \caption{Time to instantiate pods vs.\ time to pull pod images for the different \neutran components.}
    \label{fig:pod-times}
\end{figure} 

Similarly, in Figure~\ref{fig:pod-times}, we show the time taken by \neutran to instantiate each pod, as well as the time needed to download (i.e., pull) the pod image from the image registry to the physical machine where it will be instantiated.
The xApps images are the ones that take the least time to be pulled and instantiated, with an average pull time below $0.1$\:s.
A more diverse distribution can be seen in the pods of the core network, whose instantiation times range from $2$\:s to $11$\:s.
The pods of the base stations have an average pull time of $0.1$\:s but they take the longest to be instantiated (as also shown in Table~\ref{tab:inst-time}.
Finally, since the pods of the near-RT \gls{ric} are the largest ones---and are hosted on the \gls{osc} Nexus image registry instead of locally on \neutran cluster---they require more time to be pulled (i.e., up to $1$\:s), but are instantiated within $4$\:s.
These results show that, on average, \neutran can instantiate a fully functional cellular network in around $10$\:s.

\section{Conclusions}
\label{sec:conclusions}

We presented \neutran, a zero-touch framework to enable and automate \ran and spectrum sharing among multiple tenants. \neutran allows tenants to deploy on the fly an end-to-end \ran tailored to their needs. We introduced the \neutran framework, and described its components and functionalities, including its optimization engine rApp and virtualized infrastructure.
Optimal support of neutral host applications is obtained by modeling it as a complex optimization problem, that we show solvable well within seconds for networks of realistic scale (e.g., 50 cell sites).
We prototyped \neutran on an OpenShift cluster and an indoor RAN testbed with 4 base stations and 10 users from 3 different tenants, showing that \neutran can boost the cumulative throughput of tenants' subscribers by up to~2.18$\times$.

\footnotesize  
\bibliographystyle{IEEEtran}
\bibliography{biblio.bib}

\begin{thebibliography}{10}
\providecommand{\url}[1]{#1}
\csname url@samestyle\endcsname
\providecommand{\newblock}{\relax}
\providecommand{\bibinfo}[2]{#2}
\providecommand{\BIBentrySTDinterwordspacing}{\spaceskip=0pt\relax}
\providecommand{\BIBentryALTinterwordstretchfactor}{4}
\providecommand{\BIBentryALTinterwordspacing}{\spaceskip=\fontdimen2\font plus
\BIBentryALTinterwordstretchfactor\fontdimen3\font minus
  \fontdimen4\font\relax}
\providecommand{\BIBforeignlanguage}[2]{{%
\expandafter\ifx\csname l@#1\endcsname\relax
\typeout{** WARNING: IEEEtran.bst: No hyphenation pattern has been}%
\typeout{** loaded for the language `#1'. Using the pattern for}%
\typeout{** the default language instead.}%
\else
\language=\csname l@#1\endcsname
\fi
#2}}
\providecommand{\BIBdecl}{\relax}
\BIBdecl

\bibitem{bhushan2014network}
N.~{Bhushan}, J.~{Li}, D.~{Malladi}, R.~{Gilmore}, D.~{Brenner},
  A.~{Damnjanovic}, R.~T. {Sukhavasi}, C.~{Patel}, and S.~{Geirhofer},
  ``Network densification: The dominant theme for wireless evolution into
  {5G},'' \emph{IEEE Communications Magazine}, vol.~52, no.~2, pp. 82--89,
  February 2014.

\bibitem{wen2022private}
M.~Wen, Q.~Li, K.~J. Kim, D.~López-Pérez, O.~A. Dobre, H.~V. Poor,
  P.~Popovski, and T.~A. Tsiftsis, ``Private {5G} networks: Concepts,
  architectures, and research landscape,'' \emph{IEEE Journal of Selected
  Topics in Signal Processing}, vol.~16, no.~1, pp. 7--25, January 2022.

\bibitem{fccReport}
\BIBentryALTinterwordspacing
{Federal Communications Commission}, ``{WC Docket No. 18-89},'' DA 21-355,
  2021. [Online]. Available: \url{https://tinyurl.com/2maycpp3}
\BIBentrySTDinterwordspacing

\bibitem{oughton2018cost}
E.~J. Oughton and Z.~Frias, ``{The cost, coverage and rollout implications of
  5G infrastructure in Britain},'' \emph{Telecommunications Policy}, vol.~42,
  no.~8, pp. 636--652, 2018.

\bibitem{samdanis2016network}
K.~Samdanis, X.~Costa-Perez, and V.~Sciancalepore, ``From network sharing to
  multi-tenancy: The {5G} network slice broker,'' \emph{IEEE Communications
  Magazine}, vol.~54, no.~7, pp. 32--39, July 2016.

\bibitem{LAHTEENMAKI2021102201}
J.~Lahteenmaki, ``{The Evolution Paths of Neutral Host Businesses: Antecedents,
  Strategies, and Business Models},'' \emph{Telecommunications Policy},
  vol.~45, no.~10, pp. 1--27, November 2021.

\bibitem{alpha2021}
\BIBentryALTinterwordspacing
{Alpha Wireless Insights Blog}, ``{Analysis: The Neutral Host Model is
  Sprouting Wings},'' 2021. [Online]. Available:
  \url{https://alphawireless.com/analysis-the-neutral-host-model-is-sprouting-wings/}
\BIBentrySTDinterwordspacing

\bibitem{marketresearch}
\BIBentryALTinterwordspacing
{Market Research}, ``{Neutral Hosting Market Outlook and Forecasts 2021 –
  2028},'' 2021. [Online]. Available:
  \url{https://www.marketresearch.com/Mind-Commerce-Publishing-v3122/Neutral-Hosting-Outlook-Forecasts-14883308/}
\BIBentrySTDinterwordspacing

\bibitem{zhang2017survey}
L.~Zhang, M.~Xiao, G.~Wu, M.~Alam, Y.-C. Liang, and S.~Li, ``A survey of
  advanced techniques for spectrum sharing in {5G} networks,'' \emph{IEEE
  Wireless Communications}, vol.~24, no.~5, pp. 44--51, October 2017.

\bibitem{analysismason}
\BIBentryALTinterwordspacing
C.~Gabriel, ``The impact of {5G} and next-generation networks on mobile {OPEX}
  spending,'' Analysis Mason report, 2018. [Online]. Available:
  \url{https://tinyurl.com/yckndswz}
\BIBentrySTDinterwordspacing

\bibitem{stl2022open}
\BIBentryALTinterwordspacing
{STL Partners}, ``{Neutral Host: How Open RAN and Neutral Host paves the way
  for 5G},'' 2022. [Online]. Available:
  \url{https://stlpartners.com/articles/telco-cloud/neutral-host-how-open-ran-and-neutral-host-paves-way-5g/}
\BIBentrySTDinterwordspacing

\bibitem{palola2017field}
M.~Palola, M.~Höyhtyä, P.~Aho, M.~Mustonen, T.~Kippola, M.~Heikkilä,
  S.~Yrjölä, V.~Hartikainen, L.~Tudose, A.~Kivinen, R.~Ekman, J.~Hallio,
  J.~Paavola, M.~Mäkeläinen, and T.~Hänninen, ``Field trial of the 3.5 {GHz}
  citizens broadband radio service governed by a spectrum access system
  ({SAS}),'' in \emph{Proceedings of IEEE DySPAN}, Baltimore, MD, USA, March
  2017.

\bibitem{garciaaviles2021nuberu}
G.~Garcia-Aviles, A.~Garcia-Saavedra, M.~Gramaglia, X.~Costa-Perez, P.~Serrano,
  and A.~Banchs, ``Nuberu: Reliable {RAN} virtualization in shared platforms,''
  in \emph{Proceedings of ACM MobiCom}, New Orleans, LA, USA, October 2021.

\bibitem{liyanage2022survey}
M.~Liyanage, Q.-V. Pham, K.~Dev, S.~Bhattacharya, P.~K.~R. Maddikunta, T.~R.
  Gadekallu, and G.~Yenduri, ``A survey on zero touch network and service
  management {(ZSM)} for {5G} and beyond networks,'' \emph{Journal of Network
  and Computer Applications}, vol. 203, pp. 1--27, July 2022.

\bibitem{benzaid2020ai}
C.~Benzaid and T.~Taleb, ``{AI}-driven zero touch network and service
  management in {5G} and beyond: Challenges and research directions,''
  \emph{IEEE Network}, vol.~34, no.~2, pp. 186--194, March 2020.

\bibitem{doro2022orchestran}
S.~D'Oro, L.~Bonati, M.~Polese, and T.~Melodia, ``{OrchestRAN}: Network
  automation through orchestrated intelligence in the open {RAN},'' in
  \emph{Proceedings of IEEE INFOCOM}, London, United Kingdom, May 2022.

\bibitem{sharma2017dynamic}
S.~K. Sharma, T.~E. Bogale, L.~B. Le, S.~Chatzinotas, X.~Wang, and
  B.~Ottersten, ``Dynamic spectrum sharing in {5G} wireless networks with
  full-duplex technology: Recent advances and research challenges,'' \emph{IEEE
  Communications Surveys \& Tutorials}, vol.~20, no.~1, pp. 674--707, February
  2017.

\bibitem{gsma2021sharing}
\BIBentryALTinterwordspacing
{GSMA}, ``{Spectrum Sharing},'' GSMA Public Policy Position, June 2021.
  [Online]. Available: \url{https://tinyurl.com/5bju2k44}
\BIBentrySTDinterwordspacing

\bibitem{bonati2021intelligence}
L.~Bonati, S.~D'Oro, M.~Polese, S.~Basagni, and T.~Melodia, ``{Intelligence and
  Learning in O-RAN for Data-driven NextG Cellular Networks},'' \emph{IEEE
  Communications Magazine}, vol.~59, no.~10, pp. 21--27, October 2021.

\bibitem{kibria2017shared}
M.~G. Kibria, G.~P. Villardi, K.~Nguyen, W.-S. Liao, K.~Ishizu, and F.~Kojima,
  ``Shared spectrum access communications: A neutral host micro operator
  approach,'' \emph{IEEE Journal on Selected Areas in Communications}, vol.~35,
  no.~8, pp. 1741--1753, August 2017.

\bibitem{dipascale2020toward}
E.~Di~Pascale, H.~Ahmadi, L.~Doyle, and I.~Macaluso, ``Toward scalable
  user-deployed ultra-dense networks: Blockchain-enabled small cells as a
  service,'' \emph{IEEE Communications Magazine}, vol.~58, no.~8, pp. 82--88,
  August 2020.

\bibitem{vincenzi2017cooperation}
M.~Vincenzi, A.~Antonopoulos, E.~Kartsakli, J.~Vardakas, L.~Alonso, and
  C.~Verikoukis, ``Cooperation incentives for multi-operator {C-RAN} energy
  efficient sharing,'' in \emph{Proceedings of IEEE ICC}, Paris, France, May
  2017.

\bibitem{sarakis2021cost}
L.~Sarakis, P.~Trakadas, J.~Martrat, S.~Prior, O.~Trullols-Cruces, E.~Coronado,
  M.~Centenaro, G.~Kontopoulos, E.~Atxutegi, P.~Gkonis \emph{et~al.},
  ``Cost-efficient {5G} non-public network roll-out: The {Affordable5G}
  approach,'' in \emph{Proceedings of IEEE MeditCom}, Athens, Greece, September
  2021.

\bibitem{paolino2019compute}
M.~Paolino, G.~Carrozzo, A.~Betzler, C.~Colman-Meixner, H.~Khalili,
  S.~Siddiqui, T.~Sechkova, and D.~Simeonidou, ``Compute and network
  virtualization at the edge for {5G} smart cities neutral host
  infrastructures,'' in \emph{Proceedings of IEEE 5GWF}, Dresden, Germany,
  September 2019.

\bibitem{xiao2019multi}
Y.~Xiao, M.~Krunz, and T.~Shu, ``Multi-operator network sharing for massive
  {IoT},'' \emph{IEEE Communications Magazine}, vol.~57, no.~4, pp. 96--101,
  April 2019.

\bibitem{qian2020multi}
B.~Qian, H.~Zhou, T.~Ma, K.~Yu, Q.~Yu, and X.~Shen, ``Multi-operator spectrum
  sharing for massive {IoT} coexisting in {5G/B5G} wireless networks,''
  \emph{IEEE Journal on Selected Areas in Communications}, vol.~39, no.~3, pp.
  881--895, August 2020.

\bibitem{lin2017transparent}
Y.-D. Lin, H.-T. Chien, H.-W. Chang, C.-L. Lai, and K.-Y. Lin, ``Transparent
  {RAN} sharing of {5G} small cells and macrocells,'' \emph{IEEE Wireless
  Communications}, vol.~24, no.~6, pp. 104--111, December 2017.

\bibitem{giannone2019impact}
F.~Giannone, K.~Kondepu, H.~Gupta, F.~Civerchia, P.~Castoldi, A.~A. Franklin,
  and L.~Valcarenghi, ``Impact of virtualization technologies on virtualized
  {RAN} midhaul latency budget: A quantitative experimental evaluation,''
  \emph{IEEE Communications Letters}, vol.~23, no.~4, pp. 604--607, February
  2019.

\bibitem{kasgari2018stochastic}
A.~T.~Z. Kasgari and W.~Saad, ``{Stochastic Optimization and Control Framework
  for 5G Network Slicing with Effective Isolation},'' in \emph{Proceedings of
  IEEE CISS}, 2018, pp. 1--6.

\bibitem{wang2019reconfiguration}
G.~Wang, G.~Feng, T.~Q.~S. Quek, S.~Qin, R.~Wen, and W.~Tan, ``{Reconfiguration
  in Network Slicing—Optimizing the Profit and Performance},'' \emph{IEEE
  Transactions on Network and Service Management}, vol.~16, no.~2, pp.
  591--605, 2019.

\bibitem{bega2019deepcog}
D.~Bega, M.~Gramaglia, M.~Fiore, A.~Banchs, and X.~Costa-Perez, ``{DeepCog:
  Cognitive Network Management in Sliced 5G Networks with Deep Learning},'' in
  \emph{Proceedings of IEEE INFOCOM 2019}, Paris, France, April 29--May 2 2019,
  pp. 280--288.

\bibitem{foukas2019iris}
X.~Foukas, M.~K. Marina, and K.~Kontovasilis, ``Iris: Deep reinforcement
  learning driven shared spectrum access architecture for indoor neutral-host
  small cells,'' \emph{IEEE Journal on Selected Areas in Communications},
  vol.~37, no.~8, pp. 1820--1837, August 2019.

\bibitem{caballero2019network}
P.~Caballero, A.~Banchs, G.~De~Veciana, and X.~Costa-Pérez, ``{Network Slicing
  Games: Enabling Customization in Multi-Tenant Mobile Networks},''
  \emph{IEEE/ACM Transactions on Networking}, vol.~27, no.~2, pp. 662--675,
  April 2019.

\bibitem{caballero2017multi}
P.~Caballero, A.~Banchs, G.~de~Veciana, and X.~Costa-Pérez, ``{Multi-Tenant
  Radio Access Network Slicing: Statistical Multiplexing of Spatial Loads},''
  \emph{IEEE/ACM Transactions on Networking}, vol.~25, no.~5, pp. 3044--3058,
  October 2017.

\bibitem{doro2021coordinated}
S.~D'Oro, L.~Bonati, F.~Restuccia, and T.~Melodia, ``Coordinated {5G} network
  slicing: How constructive interference can boost network throughput,''
  \emph{IEEE/ACM Transactions on Networking}, vol.~29, no.~4, pp. 1881--1894,
  2021.

\bibitem{doro2020sledge}
S.~D'Oro, L.~Bonati, F.~Restuccia, M.~Polese, M.~Zorzi, and T.~Melodia,
  ``{Sl-EDGE}: Network slicing at the edge,'' in \emph{Proceedings of ACM
  Mobihoc}, Virtual Event, October 2020.

\bibitem{puligheddu2023semoran}
C.~Puligheddu, J.~Ashdown, C.~F. Chiasserini, and F.~Restuccia, ``{SEM-O-RAN:
  Semantic and Flexible O-RAN Slicing for NextG Edge-Assisted Mobile
  Systems},'' in \emph{Proceedings of IEEE INFOCOM 2023 (preprint available as
  arXiv:2212.11853 [cs.NI])}, New York Area, NJ, USA, May 2023.

\bibitem{baldesi2022charm}
L.~Baldesi, F.~Restuccia, and T.~Melodia, ``{ChARM}: {NextG} spectrum sharing
  through data-driven real-time {O-RAN} dynamic control,'' in \emph{Proceedings
  of IEEE INFOCOM}, London, United Kingdom, May 2022.

\bibitem{bega2017optimising}
D.~Bega, M.~Gramaglia, A.~Banchs, V.~Sciancalepore, K.~Samdanis, and
  X.~Costa-Perez, ``{Optimising 5G Infrastructure Markets: The Business of
  Network Slicing},'' in \emph{Proceedings of IEEE INFOCOM}, Atlanta, GA, USA,
  May 2017, pp. 1--9.

\bibitem{sciancalepore2017traffic}
V.~Sciancalepore, K.~Samdanis, X.~Costa-Perez, D.~Bega, M.~Gramaglia, and
  A.~Banchs, ``{Mobile Traffic Forecasting for Maximizing 5G Network Slicing
  Resource Utilization},'' in \emph{Proceedings of IEEE INFOCOM}, Atlanta, GA,
  USA, May 2017, pp. 1--9.

\bibitem{caballero2018netowrk}
P.~Caballero, A.~Banchs, G.~de~Veciana, X.~Costa-Pérez, and A.~Azcorra,
  ``Network slicing for guaranteed rate services: Admission control and
  resource allocation games,'' \emph{IEEE Transactions on Wireless
  Communications}, vol.~17, no.~10, pp. 6419--6432, October 2018.

\bibitem{schmidt2019flexvran}
R.~Schmidt, C.-Y. Chang, and N.~Nikaein, ``{FlexVRAN}: A flexible controller
  for virtualized {RAN} over heterogeneous deployments,'' in \emph{Proceedings
  of IEEE ICC}, Shanghai, China, May 2019.

\bibitem{moorthy2022oswireless}
S.~K. Moorthy, Z.~Guan, N.~Mastronarde, E.~S. Bentley, and M.~Medley,
  ``{OSWireless}: Enhancing automation for optimizing intent-driven
  software-defined wireless networks,'' in \emph{Proceedings of IEEE MASS},
  Denver, CO, USA, October 2022.

\bibitem{foukas2017orion}
X.~Foukas, M.~K. Marina, and K.~Kontovasilis, ``Orion: {RAN} slicing for a
  flexible and cost-effective multi-service mobile network architecture,'' in
  \emph{Proceedings of ACM MobiCom}, Snowbird, UT, USA, October 2017.

\bibitem{neutral2018}
\BIBentryALTinterwordspacing
X.~Foukas, F.~Sardis, F.~Foster, M.~K. Marina, M.~A. Lema, and M.~Dohler,
  ``Experience building a prototype 5g testbed,'' in \emph{Proceedings of the
  Workshop on Experimentation and Measurements in 5G}, ser. EM-5G'18.\hskip 1em
  plus 0.5em minus 0.4em\relax New York, NY, USA: Association for Computing
  Machinery, 2018, p. 13–18. [Online]. Available:
  \url{https://doi.org/10.1145/3286680.3286683}
\BIBentrySTDinterwordspacing

\bibitem{oran-wg1-arch-spec}
{O-RAN Working Group 1}, ``{O-RAN} architecture description 5.00,''
  O-RAN.WG1.O-RAN-Architecture-Description-v05.00 Technical Specification, July
  2021.

\bibitem{doro2022dapps}
S.~D'Oro, M.~Polese, L.~Bonati, H.~Cheng, and T.~Melodia, ``{dApps: Distributed
  Applications for Real-Time Inference and Control in O-RAN},'' \emph{IEEE
  Communications Magazine}, vol.~60, no.~11, pp. 52--58, November 2022.

\bibitem{polese2023understanding}
M.~Polese, L.~Bonati, S.~D'Oro, S.~Basagni, and T.~Melodia, ``{Understanding
  O-RAN: Architecture, Interfaces, Algorithms, Security, and Research
  Challenges},'' \emph{IEEE Communications Surveys \& Tutorials}, vol.~25,
  no.~2, pp. 1376--1411, January 2023.

\bibitem{openshift}
{Red Hat}. {OpenShift}. \url{https://tinyurl.com/382mww6d}.

\bibitem{kubernetes}
{Linux Foundation}. {Kubernetes}. \url{https://kubernetes.io}.

\bibitem{huang2016consensus}
K.~Huang and N.~D. Sidiropoulos, ``{Consensus-ADMM} for general quadratically
  constrained quadratic programming,'' \emph{IEEE Transactions on Signal
  Processing}, vol.~64, no.~20, pp. 5297--5310, July 2016.

\bibitem{androulakis2008quadratic}
P.~Androulakis and P.~M. Pardalos, ``Quadratic integer programming: Complexity
  and equivalent forms,'' \emph{Encyclopedia of Optimization. Berlin, Germany:
  Springer}, 2008.

\bibitem{boyd1997semidefinite}
S.~Boyd and L.~Vandenberghe, ``Semidefinite programming relaxations of
  non-convex problems in control and combinatorial optimization,'' in
  \emph{Communications, Computation, Control, and Signal Processing}.\hskip 1em
  plus 0.5em minus 0.4em\relax Springer, 1997, pp. 279--287.

\bibitem{sherali2013reformulation}
H.~D. Sherali and W.~P. Adams, \emph{A Reformulation-linearization Technique
  for Solving Discrete and Continuous Nonconvex Problems}.\hskip 1em plus 0.5em
  minus 0.4em\relax Springer Science \& Business Media, 2013, vol.~31.

\bibitem{anstreicher2009compare}
K.~M. Anstreicher, ``Semidefinite programming versus the
  reformulation-linearization technique for nonconvex quadratically constrained
  quadratic programming,'' \emph{Journal of Global Optimization}, vol.~43,
  no.~2, pp. 471--484, March 2009.

\bibitem{basu2017large}
K.~Basu, A.~Saha, and S.~Chatterjee, ``Large-scale quadratically constrained
  quadratic program via low-discrepancy sequences,'' \emph{Advances in Neural
  Information Processing Systems}, vol.~30, 2017.

\bibitem{3gpp.38.211}
\BIBentryALTinterwordspacing
3GPP, ``{NR}; physical channels and modulation,'' 3rd Generation Partnership
  Project ({3GPP}), Technical Specification (TS) 38.211, 06 2022, version
  17.2.0. [Online]. Available: \url{http://www.3gpp.org/DynaReport/38211.htm}
\BIBentrySTDinterwordspacing

\bibitem{argocd_website}
{Argo Project}. {Argo CD}. \url{https://argoproj.github.io/cd}.

\bibitem{tekton_website}
{The Linux Foundation}. {Tekton}. \url{https://tekton.dev}.

\bibitem{open5gs_website}
{Open5GS}. \url{https://open5gs.org}.

\bibitem{bonati2022openrangym-pawr}
L.~Bonati, M.~Polese, S.~D'Oro, S.~Basagni, and T.~Melodia, ``{OpenRAN Gym:
  AI/ML Development, Data Collection, and Testing for O-RAN on PAWR
  Platforms},'' \emph{Computer Networks}, pp. 1--12, November 2022.

\bibitem{bonati2021scope}
L.~Bonati, S.~D'Oro, S.~Basagni, and T.~Melodia, ``{SCOPE}: An open and
  softwarized prototyping platform for {NextG} systems,'' in \emph{Proceedings
  of ACM MobiSys}, Virtual, June 2021.

\bibitem{srsran_website}
{Software Radio Systems}. {srsRAN}. \url{https://www.srsran.com}.

\end{thebibliography}

\vspace{-0.9cm}

\begin{IEEEbiography}
[{\includegraphics[width=1in,height=1.25in,keepaspectratio]{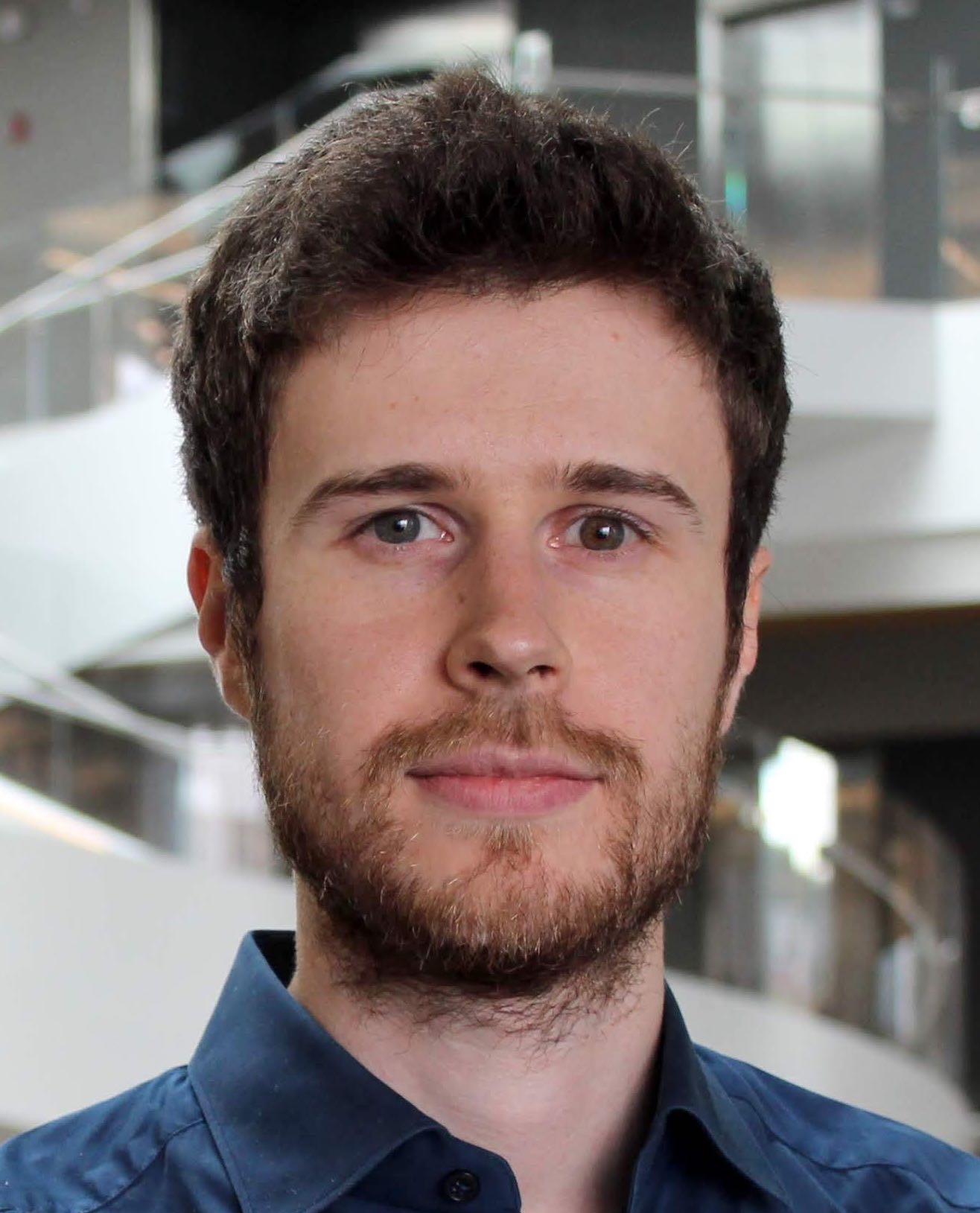}}]{Leonardo Bonati}
is an Associate Research Scientist at the Institute for the Wireless Internet of Things, Northeastern University, Boston, MA. He received the Ph.D. degree in Computer Engineering from Northeastern University in 2022. His main research focuses on softwarized approaches for the Open Radio Access Network (RAN) of the next generation of cellular networks, on O-RAN-managed networks, and on network automation and orchestration. He served on the technical program committee of the ACM Workshop on Wireless Network Testbeds, Experimental evaluation \& Characterization, and as guest editor of the special issue of Elsevier's Computer Networks journal on Advances in Experimental Wireless Platforms and Systems.
\end{IEEEbiography}

\vspace{-0.9cm}

\begin{IEEEbiography}
[{\includegraphics[width=1in,height=1.25in,keepaspectratio]{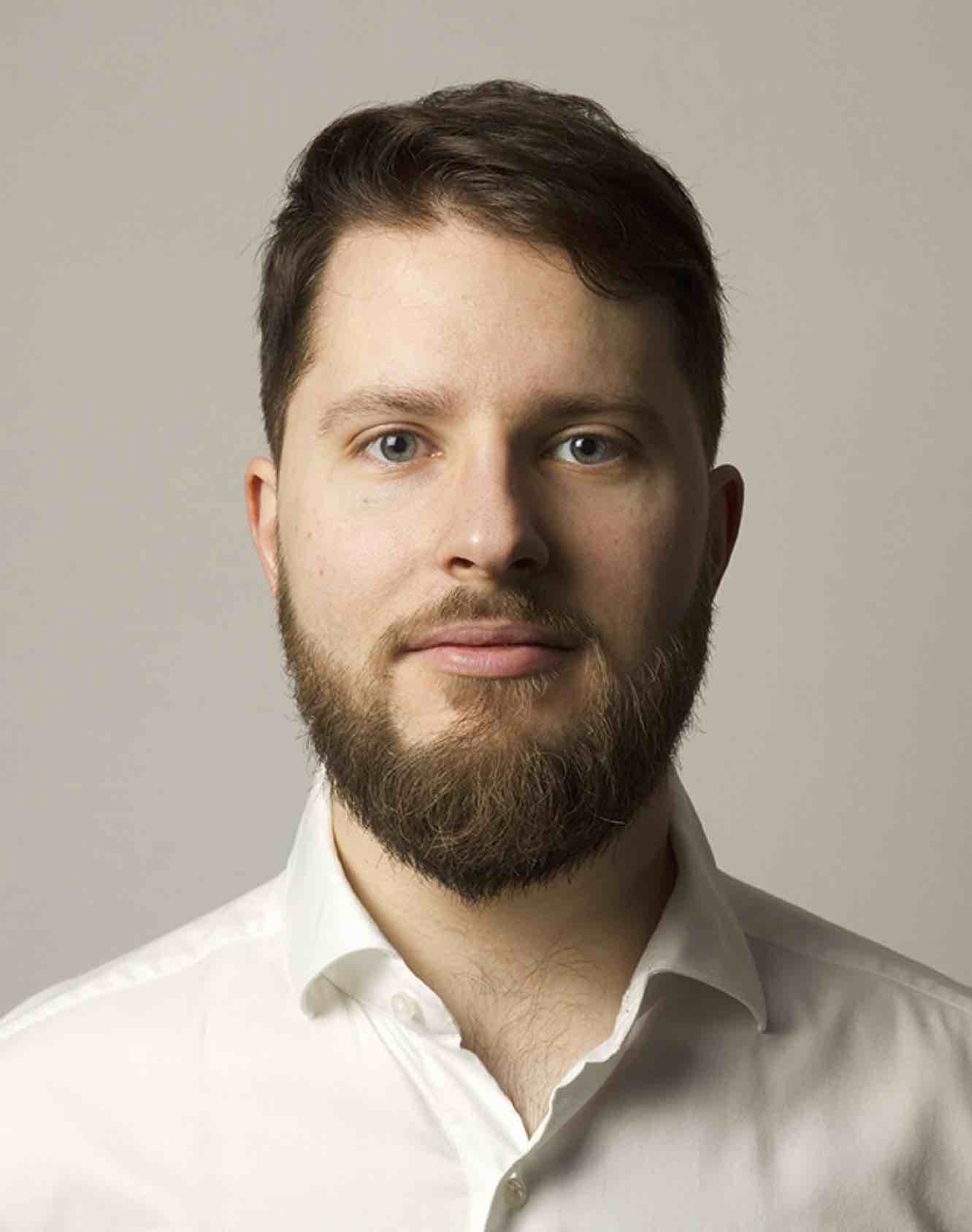}}]{Michele Polese} is a Principal Research Scientist at the Institute for the Wireless Internet of Things, Northeastern University, Boston, since March 2020. He received his Ph.D. at the Department of Information Engineering of the University of Padova in 2020. He also was an adjunct professor and postdoctoral researcher in 2019/2020 at the University of Padova, and a part-time lecturer in Fall 2020 and 2021 at Northeastern University. 
%
His research interests are in the analysis and development of protocols and architectures for future generations of cellular networks (5G and beyond).
He has contributed to O-RAN technical specifications and submitted responses to multiple FCC and NTIA notice of inquiry and requests for comments.
He received several best paper awards, is serving as TPC co-chair for WNS3 2021-2022, as an Associate Technical Editor for the IEEE Communications Magazine, and has organized the Open 5G Forum in Fall 2021.
\end{IEEEbiography}


\begin{IEEEbiography}
[{\includegraphics[width=1in,height=1.25in,keepaspectratio]{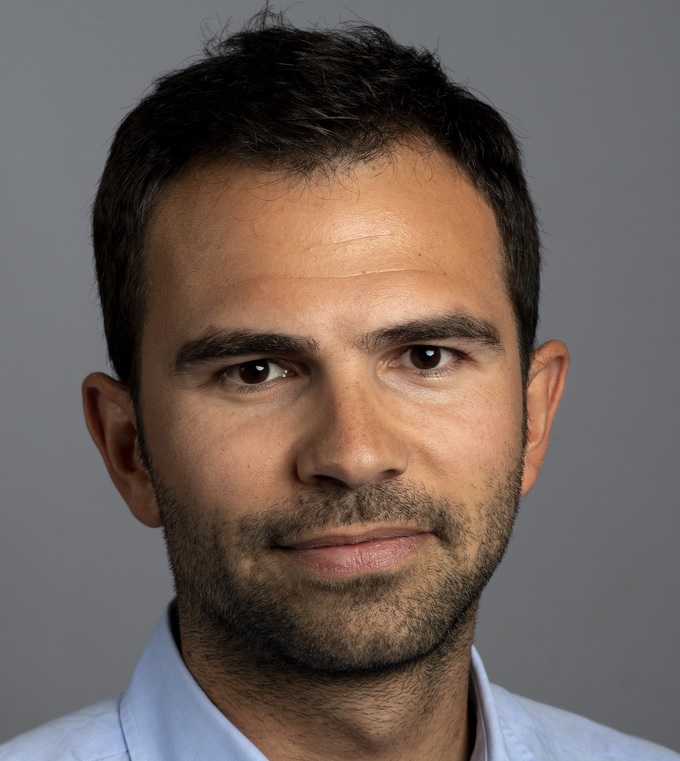}}]{Salvatore D'Oro}
is a Research Assistant Professor at Northeastern University. He received his Ph.D. degree from the University of Catania in 2015. Salvatore is an area editor of Elsevier Computer Communications journal and serves on the Technical Program Committee (TPC) of multiple conferences and workshops such as IEEE INFOCOM, IEEE CCNC, IEEE ICC and IFIP Networking. He is one of the contributors to OpenRAN Gym, the first open-source research platform for AI/ML applications in the Open RAN. Dr. D'Oro's research interests include optimization, artificial intelligence, security, network slicing and their applications to 5G networks and beyond, with specific focus on Open RAN systems.
\end{IEEEbiography}


\begin{IEEEbiography}
[{\includegraphics[width=1in,height=1.25in,keepaspectratio]{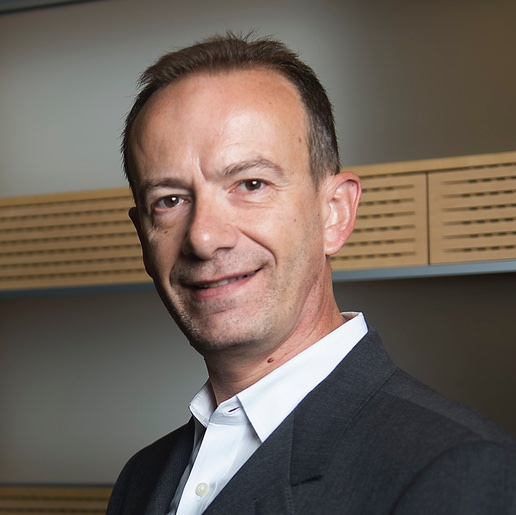}}]{Stefano Basagni}
is with the Institute for the Wireless Internet of Things and a professor at the ECE Department at Northeastern University, in Boston, MA. He holds a Ph.D.\ in electrical engineering from the University of Texas at Dallas (2001) and a Ph.D.\ in computer science from the University of Milano, Italy (1998). Dr. Basagni's current interests concern research and implementation aspects of mobile networks and wireless communications systems, wireless sensor networking for IoT (underwater, aerial and terrestrial), and definition and performance evaluation of network protocols.
Dr.\ Basagni has published over twelve dozen of highly cited, refereed technical papers and book chapters. His h-index is currently 50 (July 2023). He is also co-editor of three books. Dr. Basagni served as a guest editor of multiple international ACM/IEEE, Wiley and Elsevier journals. He has been the TPC co-chair of international conferences. He is a distinguished scientist of the ACM, a senior member of the IEEE, and a member of CUR (Council for Undergraduate Education).
\end{IEEEbiography}


\begin{IEEEbiography}
[{\includegraphics[width=1in,height=1.25in,keepaspectratio]{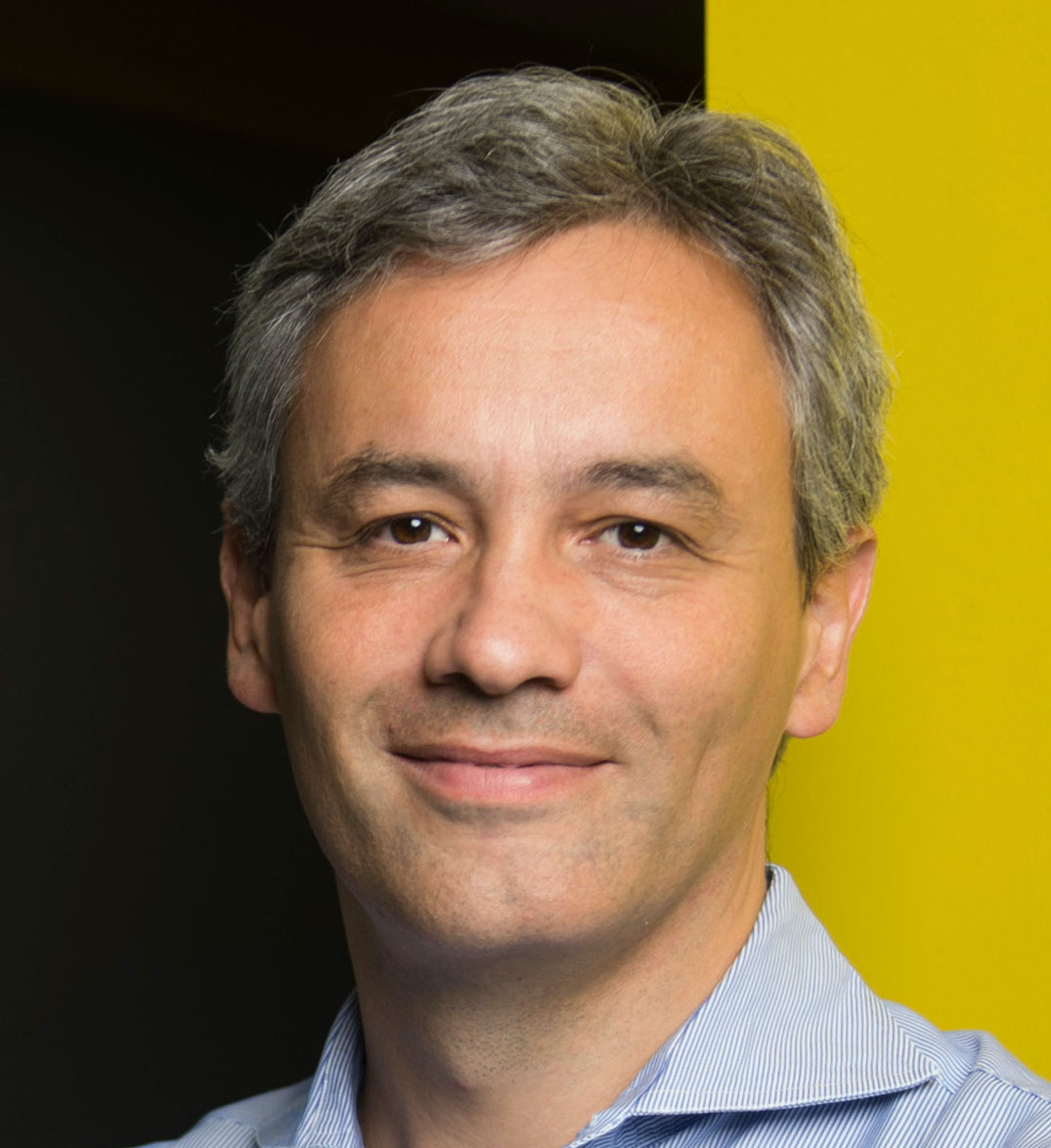}}]{Tommaso Melodia}
is the William Lincoln Smith Chair Professor with the Department of Electrical and Computer Engineering at Northeastern University in Boston. He is also the Founding Director of the Institute for the Wireless Internet of Things and the Director of Research for the PAWR Project Office. He received his Ph.D. in Electrical and Computer Engineering from the Georgia Institute of Technology in 2007. He is a recipient of the National Science Foundation CAREER award. Prof. Melodia has served as Associate Editor of IEEE Transactions on Wireless Communications, IEEE Transactions on Mobile Computing, Elsevier Computer Networks, among others. He has served as Technical Program Committee Chair for IEEE INFOCOM 2018, General Chair for IEEE SECON 2019, ACM Nanocom 2019, and ACM WUWnet 2014. Prof. Melodia is the Director of Research for the Platforms for Advanced Wireless Research (PAWR) Project Office, a \$100M public-private partnership to establish 4 city-scale platforms for wireless research to advance the US wireless ecosystem in years to come. Prof. Melodia's research on modeling, optimization, and experimental evaluation of Internet-of-Things and wireless networked systems has been funded by the National Science Foundation, the Air Force Research Laboratory the Office of Naval Research, DARPA, and the Army Research Laboratory. Prof. Melodia is a Fellow of the IEEE and a Senior Member of the ACM.
\end{IEEEbiography}

\vfill

\end{document}